\documentclass[12pt,preprint]{aastex}
\slugcomment{Accepted for publication in the ApJ on March 16, 2007}
\newcommand       \e        	[1]{\times10^{#1}}
\newcommand       \msun        	{$M_{\odot}$}
\newcommand       \lsun      	{$L_{\odot}$} 
\newcommand       \zsun      	{$Z_{\odot}$} 

\newcommand	     \cm             {cm$^{-3}$}
\newcommand	     \yr              {yr$^{-1}$}
\newcommand	     \myr              {$M_{\odot}$~yr$^{-1}$}
\newcommand       \mic        	 {$\mu$m}

\newcommand        \lya		{Ly$\alpha$}

\newcommand      \mav      {$\left<m\right>$}
\newcommand      \mave      {\left<m\right>}
\newcommand      \msn      {$\left<m_{\rm SN}\right>$}

\newcommand      \rsne      {R_{\rm SN}}
\newcommand 	     \mstar      {$m_{\star}$}
\newcommand 	     \mstare      {m_{\star}}
\newcommand      \fsn      {$f_{\rm SN}$}
\newcommand      \fsne      {f_{\rm SN}}

\newcommand		\mug			{$\mu_g$}
\newcommand		\muge		{\mu_g}
\newcommand      \yz      {$\widehat Y_z$}
\newcommand      \yze      {\widehat Y_z}
\newcommand      \yd      {$\widehat Y_d$}
\newcommand      \yde      {\widehat Y_d}
\newcommand      \md     {\left<m_d\right>}
\newcommand      \misme     {\left<m_{\rm ISM}\right>}
\newcommand      \mism     {$\left<m_{\rm ISM}\right>$}
\newcommand      \jay     {J1148+5251}

\newcommand     \nute       {\widetilde{\nu}}

\begin{document}
\bibliographystyle{/Users/edwek/Library/texmf/tex/latex/misc/aastex52/aas.bst}

\title{THE EVOLUTION OF DUST IN THE EARLY UNIVERSE WITH APPLICATIONS TO THE GALAXY SDSS~J1148+5251}

\author{Eli Dwek \altaffilmark {1}, Fr\'ed\'eric Galliano \altaffilmark{1}, \& Anthony P. Jones \altaffilmark{2}}

\altaffiltext{1}{Observational Cosmology Laboratory, Code 665, NASA Goddard Space Flight Center,
Greenbelt, MD 20771, e-mail: eli.dwek@nasa.gov}
\altaffiltext{2}{Institut d'Astrophysique Spatiale, F-91405 Orsay, France}

\begin{abstract}

Dusty hyperluminous galaxies in the early universe provide unique environments for studying the role of massive stars in the formation and destruction of dust. 
At redshifts above $\sim 6$, when the universe was less than $\sim 1$ Gyr old, dust could have only condensed in the explosive ejecta of Type~II supernovae (SNe), since most of the progenitors of the AGB stars, the major alternative source of interstellar dust, did not have time to evolve off the main sequence. In this paper we present analytical models for the evolution of the gas, dust, and metals in high redshift galaxies, with a special application to SDSS~J1148+5251 (hereafter \jay), a hyperluminous quasar at $z = 6.4$.
Ignoring accretion of interstellar matter onto grains, the main free model parameters are the dust yield in SNe, and the grain destruction efficiency by supernova remnants.
We find that an average supernova must condense at least 1~\msun\ of dust to account for the observed dust mass in \jay. Theoretically, this large yield can only be attainable if stars are formed with a top heavy initial mass function. Observationally, it is in excess of the largest dust yield of $\lesssim 0.02$~\msun\ found thus far in the ejecta of any SN. 
If future observations find this to be a typical supernova dust yield, then additional processes, such as accretion onto preexisting grains, or condensation around the AGN will need to be invoked to account for the large amount of dust in this and similar objects. The galaxy's star formation history is still uncertain, and current observations of the gas, metal, and dust contents of \jay\ can be reproduced by either an intensive and short burst of star formation ($\psi \gtrsim 10^3$~\myr) with a duration of $\lesssim~10^8$~yr, or a much lower star formation rate ($\psi \approx 100$~\myr) occurring over the lifetime of the  galaxy. Analysis of the spectral energy distribution of \jay\ suggests an AGN luminosity of $\sim 7\times10^{13}$~\lsun, requiring the early formation of a supermassive black hole of $\sim 10^9$~\msun.
\end{abstract}

 \section{INTRODUCTION}

Quasars and galaxies detected at redshifts $z \gtrsim 6$ and observed at far infrared (IR) and submillimeter wavelengths \citep{hughes98, bertoldi03a, robson04, beelen06} exhibit luminosities in excess of $\sim 10^{13}$~\lsun, and inferred dust masses and star formation rates in excess of $\sim  10^8$~\msun\ and of $\sim 10^3$~\myr, respectively \citep{bertoldi03a}. For comparison, the Milky Way contains only about $5\times 10^7$~\msun\ of dust, for an assumed gas mass of $5\times 10^9$~\msun\ and a dust-to-gas mass ratio of $\sim 0.01$. Some of these high-$z$ galaxies are younger than $\sim 1$~Gyr, strongly suggesting that the dust must have condensed in core collapse supernovae which inject their nucleosynthetic products and newly-condensed dust into the interstellar medium (ISM) relatively promptly after their formation. In contrast, AGB stars, the other main sources of interstellar dust, inject their dust products into the ISM only after their progenitor stars have evolved off the main sequence, resulting in a $\gtrsim 500$~Myr delay time since their formation \citep{dwek98}. This delayed injection of AGB-condensed dust has been suggested by \cite{dwek05c} and \cite{galliano07} as a possible explanation for the observed trend of the IR emission from polycyclic aromatic hydrocarbon (PAH) molecules with galaxies' metallicity  \citep{engelbracht05, madden06}. 

Supernovae (SNe) play a dual role in the evolution of interstellar dust. On the one hand, they are potentially the most important source of interstellar dust. If all the refractory elements precipitate with a 100\% efficiency from the gas, a typical 25~\msun\ SN can form about 1~\msun\ of dust \citep{woosley95, nomoto06}. On the other hand, their expanding blast waves sweep up interstellar dust grains, destroying them by thermal sputtering and evaporative grain-grain collisions \citep{jones96, jones04}. The presence of the large quantities of interstellar dust in these youthful galaxies therefore provides a unique opportunity for studying the dual role of SNe in the formation and destruction of dust. 

The importance of SN condensates as important contributors to the abundance of interstellar dust was previously recognized by many authors \citep{dwek80b, eales96, dwek98, tielens98, edmunds01, morgan03}. In particular \cite{eales96}, \cite{morgan03}, and \cite{maiolino04a} recognised the importance of supernova for the production of the massive amounts of dust observed in early galaxies. 

In this paper we focus on the evolution of dust in galaxies that are so youthful that AGB stars could not have contributed to the reservoir of dust in these objects. This places the burden of dust production solely on supernovae, which in the remnant phase of their evolution are also efficient agents of grain destruction. The equations for the evolution of dust are therefore greatly simplified, since: (a) they can be formulated using the instantaneous recycling approximation, which assumes that stars return their ejecta back into the interstellar medium promptly after their formation; and (b) massive core collapse supernovae are the only sources of thermally condensed dust. For the purpose of this analysis we have ignored the possible accretion of refractory elements onto preexisting dust in the dense phases of the ISM. Consequently, the models developed here allow for the detailed analysis of the dependence of the evolution of dust mass on the the dust yield in supernovae and the grain destruction efficiency by their remnants in the ISM. Furthermore, since SN play the dual role of producing {\it and} destroying dust, we also examine the dependence of dust evolution on the star formation rate.   

The paper is organized as follows. We first present the analytical models for the evolution of interstellar dust in high-redshift galaxies, assuming instantaneous recycling and neglecting the contribution of low-mass star to their chemical evolution. We also present analytical expressions for the efficiency of grain destruction by SNR in the ISM (\S2). The general properties of the equations, the evolution of gas mass, galactic metallicity, dust mass, the dust-to-gas mass ratio, and their dependence on supernova yields and grain destruction efficiencies are discussed in detail in \S3. In \S4 we apply the general results to the quasar \jay\ at redshift $z=6.4$, when the universe was a mere $\sim 900$~Myr old. We first summarize observations of this quasar, derive its far-IR luminosity and dust mass, discuss possible scenarios for the star formation history of this galaxy, and analyze its spectral energy distribution (SED), separating it into a starburst, an AGN, and dust emission components. The results of our paper are summarized in \S5.
 
 \section{THE EVOLUTION OF GAS AND DUST AT HIGH REDSHIFTS}

 \subsection{General Considerations}
 We define the stellar initial mass function (IMF), $\phi(m)$, so that $\phi(m) dm$, is the number of stars with masses between $m$ and $m+dm$, and normalize it to unity in the [$m_l,\ m_u$] mass interval, 
\begin{equation}
 \int_{m_l}^{m_u}\ \phi(m)\ dm =1
\end{equation}
 where $m_l$, and $m_u$ are, respectively, the lower and upper mass limits of the IMF. We define the IMF-averaged stellar mass, \mav, average mass of metals, \yz, and dust, \yd, returned by massive stars to the ISM as:
\begin{eqnarray}
\mave& \equiv & \int_{m_l}^{m_u}\ m\ \phi(m)\ dm   \\ \nonumber
 & & \\ \nonumber
\yze& \equiv &  \int_{m_w}^{m_{sn}}\ Y_z(m)\ \phi(m)\ dm/\fsne \\ \nonumber
 & & \\ \nonumber
\yde & \equiv & \int_{m_w}^{m_{sn}}\ Y_d(m)\ \phi(m)\ dm/\fsne \\ \nonumber
\end{eqnarray}
where $m_w$ and $m_{sn}$ are, respectively, the lower and upper mass-limits of stars that become core-collapse Type II supernovae (SN),  taken to be equal to 8 and 40~\msun\ \citep{heger03}. $Y_z(m)$ is  the total mass of metals in the gas and dust that is returned by a star of mass $m$ to the ISM,  $Y_d(m)$ is the mass of dust produced by a star of mass $m$, and \fsn\ is the fraction of stars that become SN~II, given  by:
\begin{equation}
\fsne =   \int_{m_w}^{m_{sn}}\  \phi(m)\ dm \ < \ 1
\end{equation}

We define the star formation rate (SFR), $\psi(t)$, to be  the mass of stars formed per unit time, and the stellar birthrate, $B(t)$, as the number of stars formed per unit time. For a constant IMF the relation between the two is given by: 
\begin{equation}
\psi(t) = B(t)\ \mave
\end{equation}
The Type II supernova rate is given by:
\begin{equation}
\rsne(t)   =  B(t)\ \int_{m_w}^{m_{sn}}\ \phi(m)\ dm =  {\psi(t)\over \mave}\ \fsne    \equiv   {\psi(t)\over \mstare}
\end{equation}
where 
\begin{equation}
\mstare \equiv {\mave \over f_{SN}}
\end{equation}
is the mass of all stars born per SN event.

In all our calculations we will use three different functional forms for the IMF, characterized by an $m^{-\alpha}$ power law in the [$m_l, m_u$] mass interval. Table \ref{imf} lists the parameters of the different IMFs considered in this paper. The first functional form is the Salpeter IMF, and the choice of the two additional functional forms is motivated by the possibility that at high redshifts the stellar IMF may have been more heavily weighted towards high mass stars. In addition, the table lists the IMF-averaged values of some relevant quantities. In particular, the value of \mstar\ is 147~\msun\ for a Salpeter IMF, dropping down to 50~\msun\ for a top heavy IMF which has a higher fraction of SN events.

\subsection{The Evolution of the Gas}
Let $M_g$ be the mass of gas in a galaxy. For sufficiently young galaxies we can assume that stellar ejecta are instantaneously recycled back into the ISM. In this approximation, the rate of change in the ISM mass due to astration and infall and outflow of gas is given by:
\begin{eqnarray}
{dM_g\over dt} & = & -(1-R) \psi(t) +\left( {dM_g\over dt}\right)_{inf} - \left( {dM_g\over dt}\right)_{out} \label{mg_eq} \\ \nonumber
 & = &  -(1-R) \psi(t) + \left({M_{inf}\over \tau_{inf}}\right)\times \exp\left(-{t \over \tau_{inf}}\right) -
 \left({M_{out}\over \tau_{out}}\right)\times \exp\left(-{t \over \tau_{out}}\right)
\end{eqnarray}
where $\psi(t)$ is the star formation rate, $R$ is the fraction of the stellar mass that is returned to the ISM, either by SN explosions or quiescent stellar winds. The last two terms are the net increase in the mass of the ISM to infall and outflow, respectively. We assumed them to be exponential with e-folding times of $\tau_{inf}$ and $\tau_{out}$,  normalized so that the total mass of the infalling (outflowing) gas is given  by $M_{inf}$ ($M_{out}$). 

We assume that the star formation rate is proportional to the gas mass to some power~$k$: 
\begin{eqnarray}
\psi(t) & =  & \psi_0\ \left({M_g(t)\over M_0}\right)^k \\ \nonumber
 & = &  \psi_0\ \muge(t)^k
\end{eqnarray}
where $M_0$ is the total mass of the system (stars + gas) at some arbitrary time $t_0$.

We will consider two scenarios for the chemical evolution of the system: (1) a closed box model in which a galaxy does not exchange any mass with its surrounding medium, that is,  $M_{inf} = M_{out} = 0$; and (2) an infall model, in which a galaxy's mass is initially zero, and increases with  time due to the infall of metal-free gas. 

\subsubsection{Closed Box Model}
In this model the equation for $dM_g/dt$ can then be written as:
\begin{equation}
{d\muge(t)\over dt} = -(1-R)\ \left({\psi_0\over M_0}\right)\ \muge(t)^k
\label{dmugdt_cb}
\end{equation}
where $M_0 \equiv M_g(t_0=0)$, is the total mass of the system (which initially consists of only gas), $\psi_0 = \psi(t=0)$, and $\muge(t) \equiv M_g(t)/M_0$ is the gas mass fraction at time $t$. Analytical solutions are available for arbitrary values of $k$:
\begin{eqnarray}
\muge(t) & = &  \ \exp\left[ -(1-R)\ \left({\psi_0\over M_0}\right) t \right] \qquad \qquad \qquad k=1 \\ \nonumber
 & = & \left[ 1 -(1-R)(1-k)\ \left({\psi_0\over M_0}\right) t \right]^{1\over 1-k} \qquad \ \ k \neq 1
 \label{mug_cb}
\end{eqnarray}
with the initial condition that \mug\ = 1 at time $t=0$.

\subsubsection{Infall Model}
In this model the equation for $dM_g/dt$ is:
\begin{equation}
{dM_g\over dt}  =  -(1-R) \psi(t) +\left({M_{inf}\over \tau_{inf}}\right)\times \exp\left(-{t \over \tau_{inf}}\right)
\end{equation}
Since the initial mass of the system is zero, we define $M_0$ to be the total mass of the galaxy at some later time $t_0 > 0$, which, in the absence of outflows, is given by:
\begin{eqnarray}
M_0 & \equiv &  \left({M_{inf}\over \tau_{inf}}\right)\ \int_0^{t_0}\ \exp\left(-{t\over \tau_{inf}}\right)\ dt \\ \nonumber
 & = & M_{inf}\ \left[1-\exp\left(-{t_0\over \tau_{inf}}\right)\right]
\end{eqnarray}
Analytical solutions are available for $k=1$ and constant values of $M_{inf}$ and $\tau_{inf}$, or for arbitrary values of $k$ provided that $(M_{inf}/\tau_{inf}) \propto \psi(t)$. 

Defining $\muge(t) \equiv M_g(t)/M_0$,  the solution for the evolution of $\muge(t)$ for $k=1$ can be written as:
\begin{equation}
\muge(t) = \tilde \mu\ \left\{\exp\left[-{t\over\tau_{inf}}\right] - \exp\left[-(1-R){t\over \tau_0}\right]\right\}
\label{mug_inf}
\end{equation}
where $\tau_0  \equiv  M_0/\psi_0$, and
\begin{eqnarray}
\tilde \mu  & \equiv & {M_{inf} \over (1-R)\, \psi_0\, \tau_{inf} - M_0} \label{mu_tilde} \\ 
&  = &  {1 \over \left[1-\exp\left(\displaystyle-{t_0\over \tau_{inf}}\right)\right] \left[(1-R)\displaystyle\left({\tau_{inf} \over \tau_0}\right) - 1\right]} \nonumber 
\end{eqnarray}

\noindent
For a given galactic mass, $M_0$, there exist only a limited range of values for $\psi_0$ that will produce a given gas mass fraction, $\mu_g(t_0)$, at  time $t_0$, over a wide range of values for $\tau_{inf}$. The range of values for $\psi_0$ is given by:
\begin{equation}
{\ln[\mu_g(t_0)]\ M_0\over 1-R}\ \leq \ \psi_0\ t_0 \ \leq \ {x\ M_0\over 1-R}
\end{equation}
where $x$ is the solution for $[1-\exp(x)]/x = \mu_g(t_0)$. For values of $\tau_{inf}$ ranging from $\tau_{inf} \ll t_0$ to $\tau_{inf} \gg t_0$, $\psi_0$ ranges from $\sim 130$ to 270~\myr.  
\subsection{The Evolution of the Dust}
In very young galaxies with ages $\lesssim 800$~Myr, low-mass stars do not have time to evolve off the main sequence and enrich the ISM with their dusty ejecta. We will assume that any infalling gas is dust-free, and that the dust only condenses in SN~II and does not grow by accretion of metals onto grains in the ISM. The equation for the evolution of the mass of dust, $M_d$ in the  galaxy is then given by:

\begin{equation}
{dM_d(t)\over dt} = -Z_d\ \psi(t) + \yde R_{SN}(t)- {M_d(t)\over \tau_d}
\end{equation}
where $Z_d \equiv M_d/M_g$ is the dust-to-gas mass ratio, \yd\ is the average yield of dust in Type~II supernovae, $R_{SN}$ is the supernova rate in the galaxy, and $\tau_d$ is the lifetime of the dust against destruction by SN blast waves. The lifetime for grain destruction is given by (Dwek \& Scalo 1981; McKee 1987):
\begin{equation}
\tau_d = {M_d(t)\over \md R_{SN}} = {M_g(t)\over \misme R_{SN}}
\label{dust_life}
\end{equation}
where $\left<m_d\right>$ is the total mass of elements that are locked up in dust and returned by a single supernova remnant (SNR) back to the gas phase either by thermal sputtering or evaporative grain-grain collisions \citep{jones96, jones04}, and  $\misme \equiv \left<m_d\right>/Z_d$ is the effective ISM mass that is completely cleared of dust by a single SNR, more formally defined in \S2.6 below.

Using the expression for the SN rate given by eq. (5), the equation for the evolution of dust can now be written as:
\begin{equation}
{dM_d(t)\over dt} = -\left({\psi(t)\over M_g(t)}\right)\ \left[1+{\misme\over \mstare}\right]\ M_d(t) + \yde\ {\psi(t)\over \mstare}
\label{dmd_dt}
\end{equation}
Solutions for $M_d(t)$ depend on the assumed chemical evolution model.

\subsubsection{Closed Box Model}
In  this model we can change variables from $t$ to $\mu_g$ and, using the expression for $d\muge/dt$ [eq.~(\ref{dmugdt_cb})], rewrite equation (\ref{dmd_dt}) in the following form:
\begin{equation}
{dM_d\over d\mu} =  \left({\nu  \over \muge}\right)\ M_d- \yde\ {M_0\over (1-R)\ \mstare}
\label{dmd_dmu}
\end{equation}
where
\begin{equation}
\nu \equiv {\misme + \mstare \over (1-R)\ \mstare}
\label{nu_eq}
\end{equation}
The solution to eq. (\ref{dmd_dmu}) gives the evolution of dust mass (or dust mass fraction) as a function of the fractional gas mass:
\begin{eqnarray}
M_d(\muge) & = & \yde\ \left[{M_0\over \misme + R\ \mstare}\right]\ \muge \left(1-\muge^{\nu-1}\right) \label{md_mu} \label{md_eq}\\ \nonumber
\mu_d(\muge)& = &  \left[{\yde \over \misme + R\ \mstare}\right]\ \muge \left(1-\muge^{\nu-1}\right)
\end{eqnarray}
where $\mu_d \equiv M_d/M_0$, and where we assumed that $M_d = 0$ at time $t=0$, when $\muge = 1$.
Equation (\ref{md_mu}) can be rewritten to give the dust yield required to obtain a given dust-to-gas mass ratio, $Z_d$,  at a given gas mass fraction \mug:
\begin{equation}
\yde =Z_d\ \left[{\misme + R\ \mstare \over 1-\muge^{\nu-1}}\right]
\label{yd_zd}
\end{equation}

A brief glance at eq.~(\ref{md_mu}) might suggest that  the dust mass scales linearly with $M_0$, the total mass of the galaxy. However, the dependence of $M_d$ on $M_0$ is more complex, and depends on the amount of grain destruction in the galaxy. When grain destruction is efficient so that \mism $\gg R\ \mstare$, the values of $\nu \ll 1$ and $\muge^{\nu -1} \ll 1$, and eq.~(\ref{md_mu}) approaches the asymptotic value of:
\begin{equation}
M_d(\muge)=\yde\ {M_g \over \misme}
\label{md_asymp}
\end{equation}
which is independent on the initial mass of the galaxy. The dust yield required to obtain a given dust-to-gas mass ratio then becomes:
\begin{equation}
\yde= Z_d \ \misme
\label{yd_asymp}
\end{equation}
As we will see in \S4.3, these asymptotic limits are important when the total mass of the galaxy is very uncertain, but dust and gas masses are fairly well determined.   
\subsubsection{Infall Model}
In the infall model, the solution for $M_d(t)$, or equivalently $\mu_d$, cannot be simply expressed as a function of \mug(t). It has an explicit time dependence and is given by:
\begin{equation}
\mu_d(t) = \left({\yde\over \mstare}\right)\,\tilde \mu \ \left[ A\,\exp\left(-{t\over \tau_{inf}}\right)  - B\, \exp\left(-(1-R){t\over \tau_0}\right) + (B-A)\,\exp\left(-\xi\,{t\over \tau_0}\right) \right]
\label{mudt_inf}
\end{equation}
where $\tilde \mu$ is given by eq.~(\ref{mu_tilde}), $\tau_0  \equiv  M_0/\psi_0$, and 
\begin{equation}
\begin{array}{rcl}
 \xi & \equiv & 
 \displaystyle{\misme/\mstare + 1}    \\ \nonumber
A & \equiv & \left[ \xi  - \tau_0/\tau_{inf}\right]^{-1}    \\ \nonumber
B & \equiv &  \left[\xi - (1-R)\right]^{-1}  \nonumber
\end{array}    
\end{equation}

\subsection{The Evolution of the Metallicity}
The evolution of the mass of metals, $M_z(t)$, (elements heavier than helium) can be formally obtained by substituting \yz\ instead of \yd\ and by letting \mism\ $\rightarrow 0$ in equations~(\ref{nu_eq}) and (\ref{md_eq}) for $\nu$ and $M_d(t)$. 

\subsubsection{Closed Box Model}
The solution for $M_z(t)$, and $\mu_z(\muge) \equiv M_z(t)/M_0$ in this model is given by:

\begin{eqnarray}
M_z(\muge) & = & \yze\ \left({M_0\over  R\, \mstare}\right)\ \muge \left(1-\muge^{\nute-1}\right) \\ \nonumber
\mu_z(\muge) & = &  \left({\yze\over R\, \mstare}\right)\, \left(1-\muge^{\nute-1}\right)
\end{eqnarray}
where 
\begin{equation}
\nute = {1 \over (1-R)}
\end{equation}

The metallicity of the gas, $Z_g$, defined as the mass ratio of metals and the interstellar gas is then given by:
\begin{equation}
Z_g(\muge) \equiv {M_z(\muge) \over M_g(\muge)} =  \left[{\yze \over   R\ \mstare}\right]\  \left(1-\muge^{\nute-1}\right)
\end{equation}

\subsubsection{Infall Model}
Substituting \yd\ into \yz\ and setting \mism\ to 0 in the solutions of $M_d(t)$ in the infall model gives the following equation for $\mu_z(t) \equiv M_z(t)/M_0$:
\begin{equation}
\mu_z(t) = \left({\yze\over \mstare}\right)\,\tilde \mu \ \left[ A'\,\exp\left(-{t\over \tau_{inf}}\right)  - B'\, \exp\left(-(1-R){t\over \tau_0}\right) + (B'-A')\,\exp\left(-{t\over \tau_0}\right) \right]
\label{muzt_inf}
\end{equation}
where $A'=\tau_{inf}/(\tau_{inf}-\tau_0)$, and $B'=R^{-1}$.

The metallicity of the gas, $Z_g(t)$, is given by the ratio: $\mu_z(t)/\muge(t)$, where \mug(t) is given by eq.~(\ref{mug_inf}).

\subsection{The Dust to Metal Mass ratio}
An important quantity is the fraction of the mass of metals in the ISM that is locked up in dust, $f_d$, defined as:

\begin{equation}
f_d \equiv {M_d \over M_z}
\end{equation}

\noindent
For the closed box model, $f_d$ is given by:
\begin{equation}
f_d  = {\yde \over \yze}\ \left[{R\ \mstare \over \misme\ + R\ \mstare}\right] \ 
{1-\muge^{\nu-1} \over 1-\muge^{\nute-1}} 
\end{equation}
For the infall model, $f_d$ is given by the ratio $\mu_d(t)/\mu_z(t)$, given by eqs.~(\ref{mudt_inf}) and (\ref{muzt_inf}), respectively. 

\begin{figure}[htbp]
  \begin{center}
\includegraphics[width=6.0in]{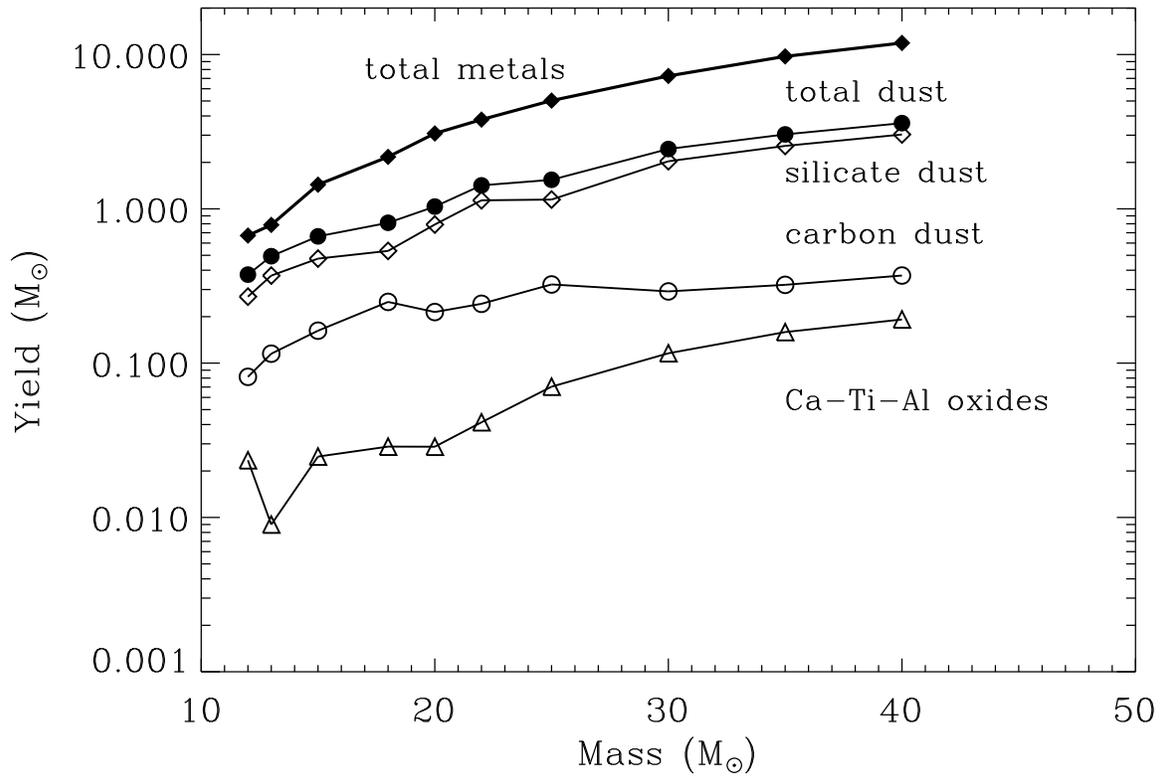}  
\end{center}
 \caption{{\footnotesize The yield of metals and dust in massive stars for $Z=Z_{\odot}$, using the yields of \cite{woosley95}. Dust yields assume that all refractory elements precipitate out of the ejecta with a condensation efficiency of unity.}}
   \label{ydyz}
\end{figure}

An upper limit to this quantity can be obtained by assuming that the dust mass is not altered in the ISM, either by grain destruction in SNRs or by accretion in molecular clouds. The value of $f_d$ is then identical in the infall or closed box model, and given by:
\begin{equation}
f_d = {\yde\over \yze} \qquad \qquad {\rm no\ grain\ destruction\ or\ accretion\ in\ the\ ISM}
\end{equation}
which is the mass fraction of metals in the SN ejecta that condensed into dust grains.
Table~\ref{dustyield} gives the values of $Y_z$, $Y_d$ for different dust compositions, the total dust yield $Y_d$, and the mass fraction of metals that form dust in SN ejecta as a function of stellar mass for two different initial metallicities of the progenitor stars. Stellar yields were taken from \cite{woosley95}. The results are also shown in Figure~\ref{ydyz} for an initial stellar metallicity of $Z_g=$~\zsun. The calculations assume that all refractory elements form dust grains with a condensation efficiency of unity. To estimate the amount of oxygen that gets incorporated into the dust we assumed that the silicate dust grains are of the form [Mg, Fe]$_2$SiO$_4$, and the Ca-Ti-Al oxides are of the form: SiO$_2$, TiO$_2$, and CaO.  

The results show that the mass fraction of metals that can form dust actually {\it decreases} with stellar mass, as more massive stars form larger amounts of oxygen which is less efficiently incorporated into dust.
The IMF-averaged mass fraction of metals that can be locked up in dust is less than $\sim 0.40$ (see Table 1). This is a strong upper limit, since these results ignore the effect of grain destruction. An observed dust mass fraction of this magnitude in a galaxy in which grain destruction is important must therefore imply that the dust mass must be enhanced by additional processes, such as the accretion of metals onto grains in the ISM.

\subsection{The Lifetime of Interstellar Dust Grains}

\subsubsection{Homogeneous Interstellar Medium}
An important input parameter governing the evolution of the dust is its lifetime against destruction by SNR. The mass of ISM gas which has been cleared of dust by a single SNR is a measure of the efficiency of this process, and is given by:
\begin{equation}
\misme = \int_{v_0}^{v_f}\ \zeta_d(v_s)\ \left|{dM\over dv_s}\right|\ dv_s
\end{equation}
where  $ \zeta_d(v_s)$ is the fraction of the mass of dust that is destroyed in an encounter with a shock wave with a velocity $v_s$, $(dM/dv_s)dv_s$ is the ISM mass that is swept up by shocks in the [$v_s,\ v_s+dv_s$] velocity range, and $v_0$ and $v_f$ are the initial and final velocities of the SNR. 

Chioffi, McKee, \& Bertschinger (1988; hereafter CMB88) presented analytical solutions for the evolution of a SNR, expanding into a uniform ISM. Initially, the remnant expands adiabatically and its evolution is described by the Sedov-Taylor solution. When the cooling time of the shocked gas becomes comparable to its dynamical timescale, the remnant evolves as a pressure-driven snowplow (PDS). For sake of our analysis we cast the standard solutions in the form of the mass of the swept-up ISM as a function of shock velocity. The evolution of the remnant mass, $M_{snr}$, in a medium with solar metallicity can then be written as:
\begin{eqnarray}
M_{snr}(M_{\odot}) & = & 400\ E_{51}^{0.86}\ n_0^{-0.28}\ v_*^{-\alpha}  \\ \nonumber
\alpha & = &
\left\{\begin{array}{lcr}
 -2.0 & {\rm for} & v_* \gtrsim 1.0 \\ \nonumber
 -1.28 & {\rm for} &  v_* \lesssim 1.0  
 \end{array}\right. \\ \nonumber
 {dM\over dv_s} & = & - \alpha\ \left({M_{snr}\over v_s}\right)
\end{eqnarray}
where $E_{51}$ is the energy of the explosion  in units of $10^{51}$~erg, $n_0$ the density of the ISM in cm$^{-3}$, and $v_* \equiv v_s/v_{\rm PDS}$, where $v_{\rm PDS}$ is the velocity of the  remnant when it transitions from the Sedov-Taylor to the PDS phase of its evolution. For a gas of solar abundances and a He/H number ratio of 0.10, $v_{\rm PDS}$ is given by (CMB88):
\begin{equation}
v_{\rm PDS} = 413\ n_0^{1/7}\ E_{51}^{1/14} \qquad \qquad {\rm km\ s}^{-1}
\end{equation}

\begin{figure}[htbp]
  \begin{center}
  \begin{tabular}{cc}
\includegraphics[width=3.1in]{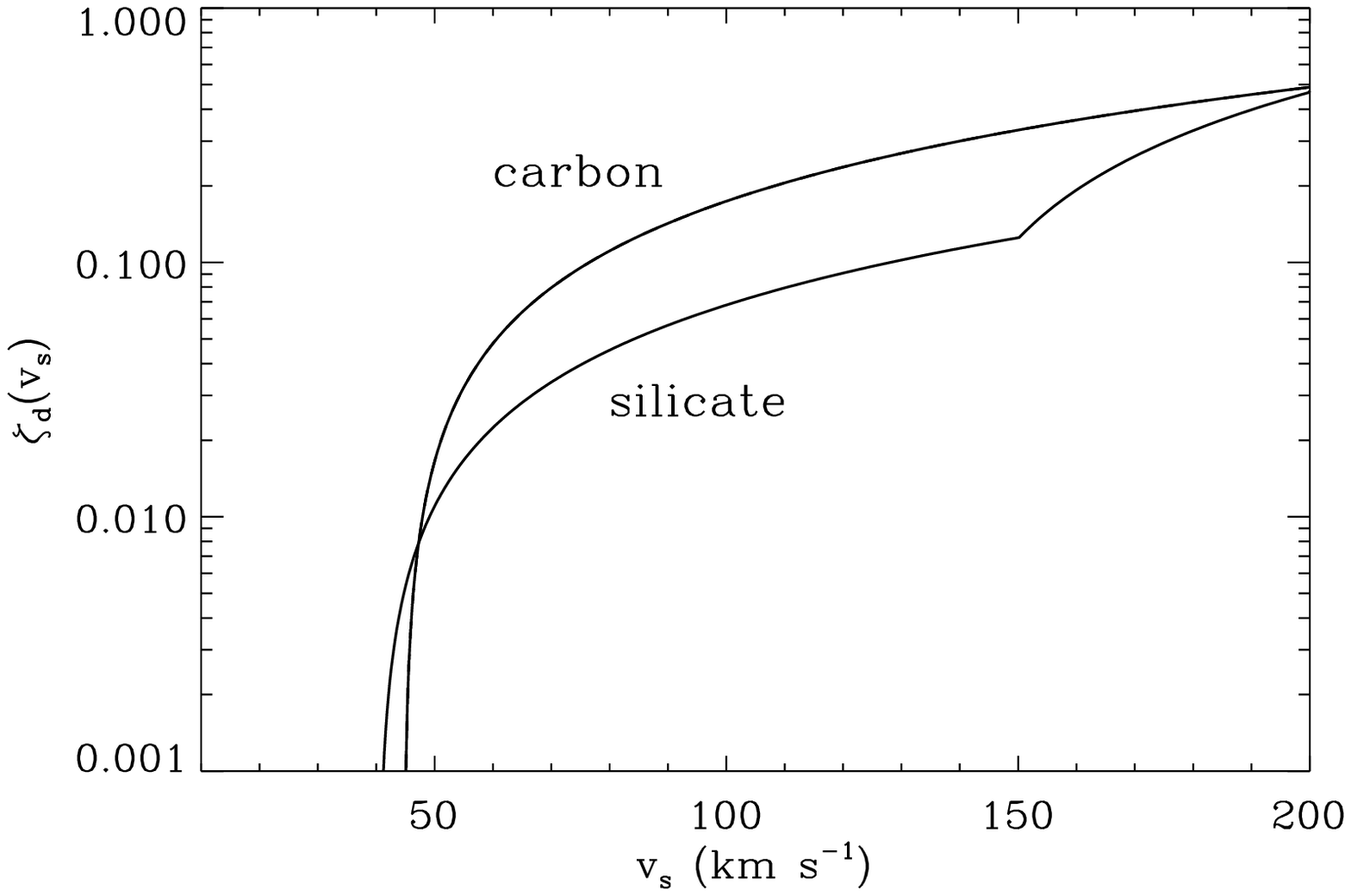}  &
\includegraphics[width=3.1in]{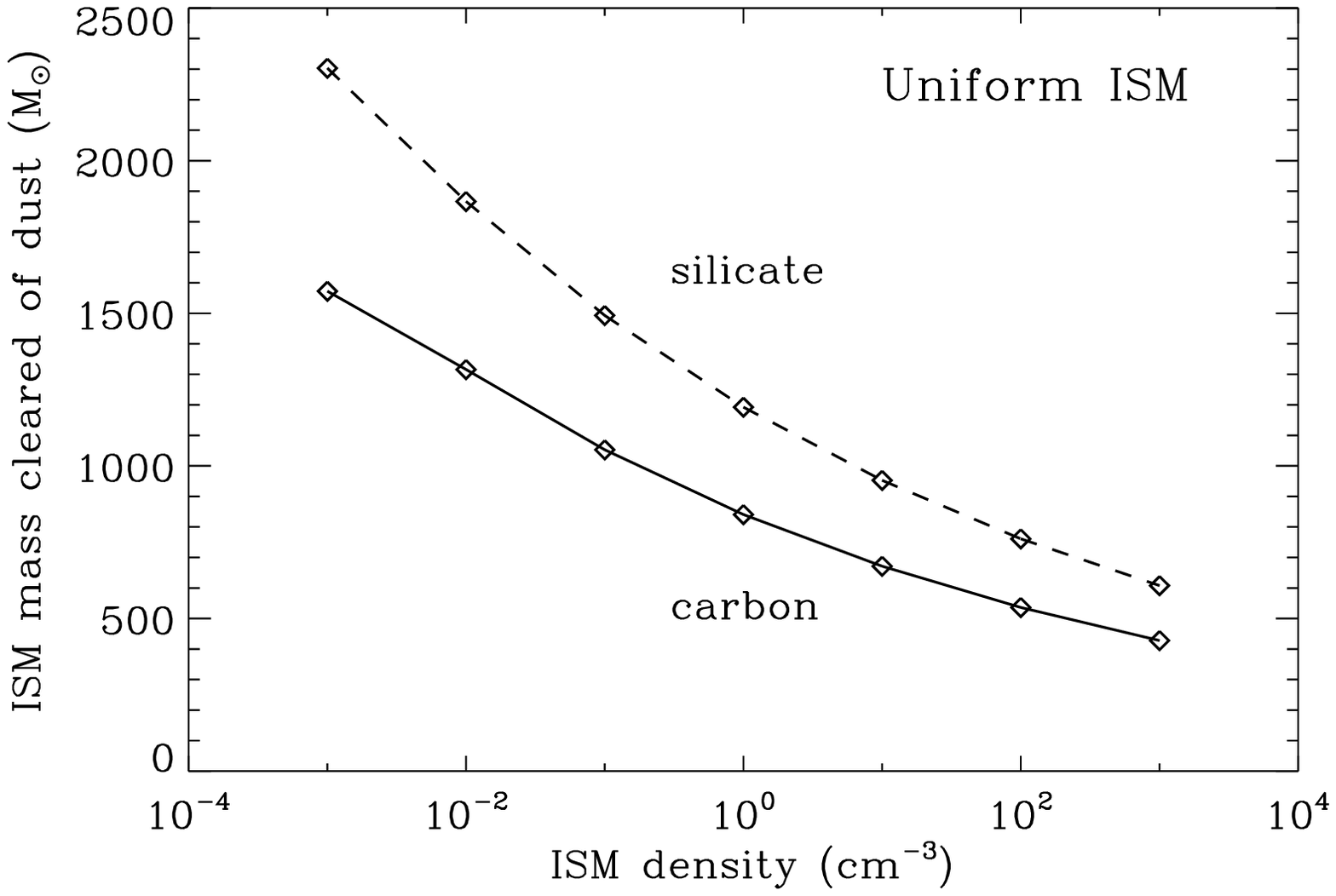} 
\end{tabular} 
\end{center}
 \caption{{\footnotesize {\bf Left panel} - the fraction of the mass of dust that returned to the gas phase by after being swept up by a shock of velocity $v_s$ as a function of shock velocity [after \cite{jones96}]. {\bf Right panel} - the mass of the ISM gas that is cleared of dust, \mism, is plotted against the density of the homogeneous ISM into which the SNR is expanding. The figure depicts the value of \mism\ for silicate (dotted line) and carbon (solid line) grains. The relation between grain lifetime and \mism\ is given in eq. (\ref{dust_life}). }}
   \label{mism}
\end{figure}

Figure~\ref{mism} depicts the velocity dependence of $\zeta_d(v_s)$ and the value of \mism\ as a function of ISM density. For values of $n_0 \approx 0.1 - 1.0$~\cm, corresponding to the average density of the Galactic ISM, we get that \mism\ $\approx 1100 - 1300$ \msun\ for an equal mix of silicate and graphite grains. For a  Galactic ISM mass of $M_g = 5\times 10^9$~\msun, and SN rate of 0.01~\yr, we get an average dust lifetime $\tau_d \approx 4\times 10^8$~yr, in good agreement with the value of $\approx 4\times 10^8$~yr ($\approx 6\times 10^8$~yr) derived by \cite{jones96} and \cite{jones04} for the average lifetime of silicate (carbon) dust in the Milky Way.

\subsubsection{Inhomogeneous Interstellar Medium}
The ISM in galaxies is inhomogeneous, and when the SFR is sufficiently high, as the case is for \jay, it is dominated by a hot and low-density gas created by the expanding SN blast waves. The multi-phase ISM is then characterized by hot ($h$), warm ($w$), and cold ($c$) phases with volume filling factors and densities of $f_i$ and $\rho_i$, with $\sum f_i = 1$ (i=\{h,w,c\}). SN blast waves  propagate predominantly through the low density intercloud medium. The shocked phases are in rough pressure equilibrium so that the velocity of the shock propagating through, say, the warm phase is related to its velocity in the hot phase by: $\rho_h v_h^2 \approx \rho_w v_w^2$.  The value of \mism\ for a 3-phase ISM can be written as:
\begin{equation}
\misme = \sum_{i=\{h,w,c\}} \ f_i\ \int_{v_0}^{v_h}\ f_d\left(v_s/\chi_i \right)\ \left|{dM\over dv_s}\right|\ dv_s
\end{equation}
where $\chi$ is the density contrast between the phase "$i$" and the dominant hot phase into which the remnant is expanding. Since the density contrast between the warm or cold phases to that of the hot ISM can be quite large, the shocks propagating through these clouds are quite ineffective in destroying the dust (Jones 2004). For a density of $n_h \approx 3\times 10^{-3}$~\cm\ and a density contrast $\chi \gtrsim 10^3 (\gtrsim 10^6)$ between the warm (cold) and hot phases, grain destruction in these phases is neglible. 

However, the warm and cold phases are ultimately cycled through the hot phase of the ISM by cloud evaporation, cloud crushing by shocks, or cloud disruption by star formation.  Injected into a hot ($\approx 10^6$~K) ISM, a dust grain of radius $a$ will be destroyed by thermal sputtering on a time scale of \citep{dwek96a, jones04}:
\begin{equation}
\Delta t_{sput} \approx 10^6\ {a(\mu {\rm m})\over n_H({\rm cm}^3)} \qquad {\rm yr}
\end{equation}
A grain of radius $a = 0.1$~\mic\ will therefore survive for a period of $\Delta t_{sput} \approx 3\times 10^7$~yr. In high redshift galaxies with total gas masses $\gtrsim 10^{10}$~\msun, and star formation rates in excess of $10^3$~\myr,  the time scale for the disruption of the cold molecular clouds by star formation, $\approx M_g/\psi \gtrsim 10^7$~yr, comparable to $\Delta t_{sput}$. The effective lifetime of the dust in the 3-phase ISM of these objects is therefore determined by $\Delta t_{sput}$. Using eq.~(\ref{dust_life}) this lifetime can be expressed in terms of \mism, giving $\misme \approx 50$~\msun, for $M_g = 10^{10}$~\msun, $\psi=10^3$~\myr, and \mstar = 150~\msun. 

To keep the model results most general, we will consider the grain destruction efficiency as an unknown, and adopt \mism\ as a free parameter of the model ranging from \mism\ = 0 (no grain destruction) to a value of \mism\ = 1000~\msun.   

\section{GENERAL RESULTS}

Figures 3 - 6 depict the evolution of various quantities as a function of fractional gas mass, \mug, for closed-box and infall models as a function of time. Results are presented for different values of \mism, ranging from 0 to $10^3$~\msun, corresponding to the range of uncertainty in the lifetime of the interstellar dust grains. The value of $R$, the IMF-averaged fraction of the stellar mass that is returned to the ISM over the stellar lifetime, is taken to be 0.50. Two different functional forms were used for the IMF: a Salpeter and a top-heavy IMF. The IMF parameters and the  values of relevant IMF-averaged quantities are given in Table~\ref{imf}. The values of the IMF-averaged dust and gas yields, \yd\  and \yz\  respectively, are given in Table~\ref{imf}.  

\subsection{The Evolution of the Gas}

\begin{figure}[htbp]
  \begin{center}
  \begin{tabular}{cc}
\includegraphics[width=3.0in]{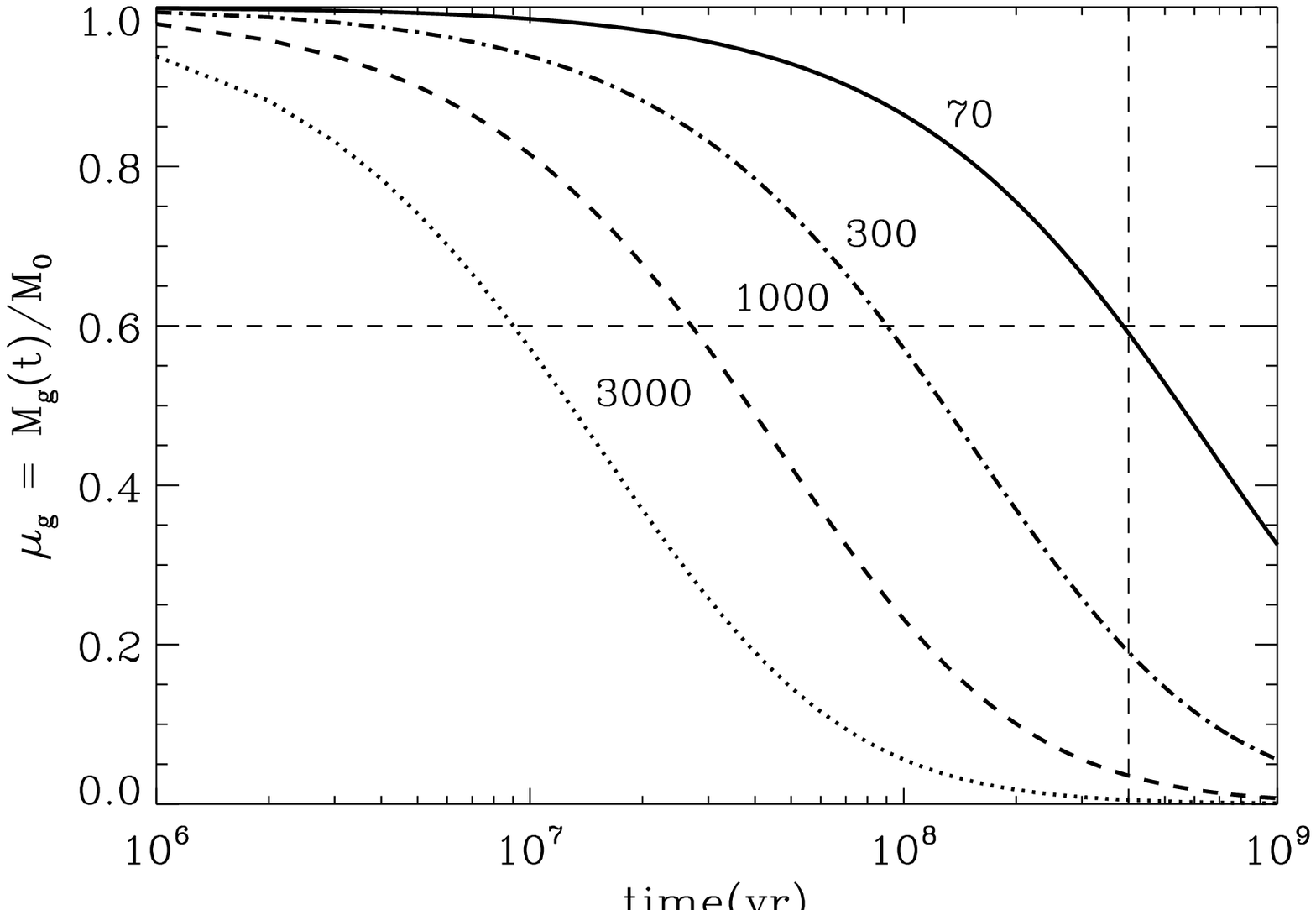} &
\includegraphics[width=3.0in]{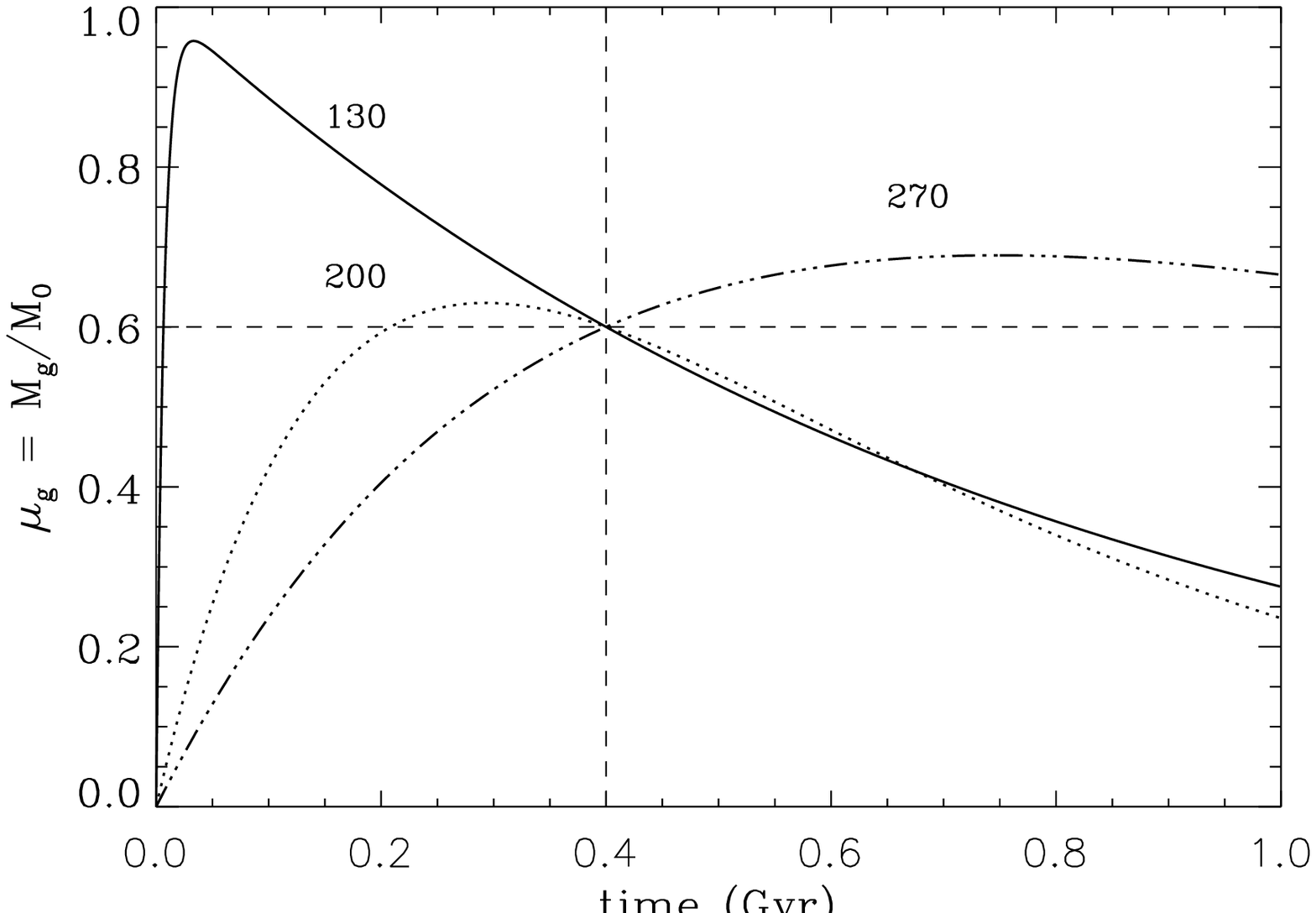}
\end{tabular} 
\end{center}
 \caption{{\footnotesize The evolution of the gas mass fraction, $\mu_g(t)$, as a a function of time for the closed box (left panel) and the infall model (right panel). The various curves are labeled by the current SFR, $\psi(t_0=400~Myr)$ for the closed box model, and by $\psi_0$ for the infall model. Both quantities are given in units of \myr. The dashed lines depict the values of the adopted mass fraction, $\muge = 0.60$, and the age of the quasar, $t = 400$~Myr, at $z$ = 6.4. The figure is discussed in further detail in \S3.1 of the text.} }
    \label{mutime}
\end{figure}

Figure~\ref{mutime} depicts the evolution of the gas mass fraction as a function of time for the closed box and infall models. In the closed box model (left panel), the initial gas mass fraction is equal to 1, decreasing as the ISM gas is converted into stars. The calculations were performed for a SFR law $\psi(t)= \psi_0 (M_g(t)/M_0)^k$ with $M_0 = 5\times 10^{10}$~\msun, and $k=1.5$. The curves are labeled by the value of $\psi(t_0)$ at time $t_0=400$~Myr, which is our adopted age of \jay\ (see \S4.1). 

In the infall model (right panel), described by eq.~(\ref{mug_inf}), the initial gas mass fraction is zero. It first increases in time as the galaxy accretes mass from its surrounding, but decreases later on, when star formation consumes gas at a higher rate than its rate of replenishment by infall. The curves are labeled by $\psi_0$, which is related to the current SFR by $\psi_0=\mu_g(t_0)\, \psi(t_0)$. For each value of $\psi_0$, eq.~(\ref{mug_inf}) was solved for the value of $\tau_{inf}$ that produced the adopted values of $M_0$ and $\mu_g$ at the epoch of $t_0$ = 400~Myr.

Figure \ref{mutime} is important for reconstructing the star formation history of a galaxy from current observations of the fractional gas mass and the star formation rate. The figure is used in \S4.5 to construct possible star formation scenarios for \jay.

\subsection{The Evolution of the Dust and Metals}

\begin{figure}[h]
  \begin{center}
  \begin{tabular}{cc}
\includegraphics[width=3.0in]{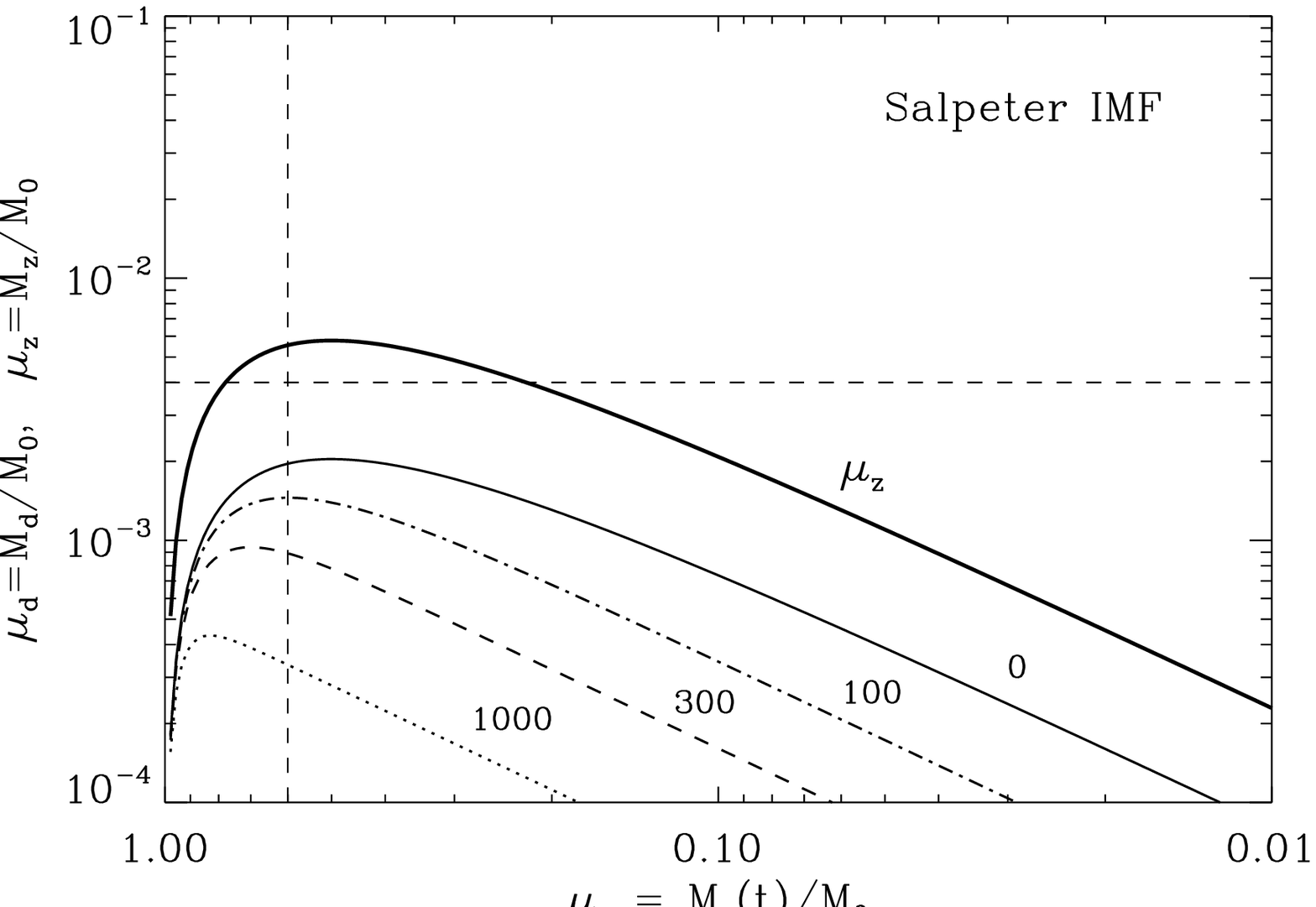}  &
\includegraphics[width=3.0in]{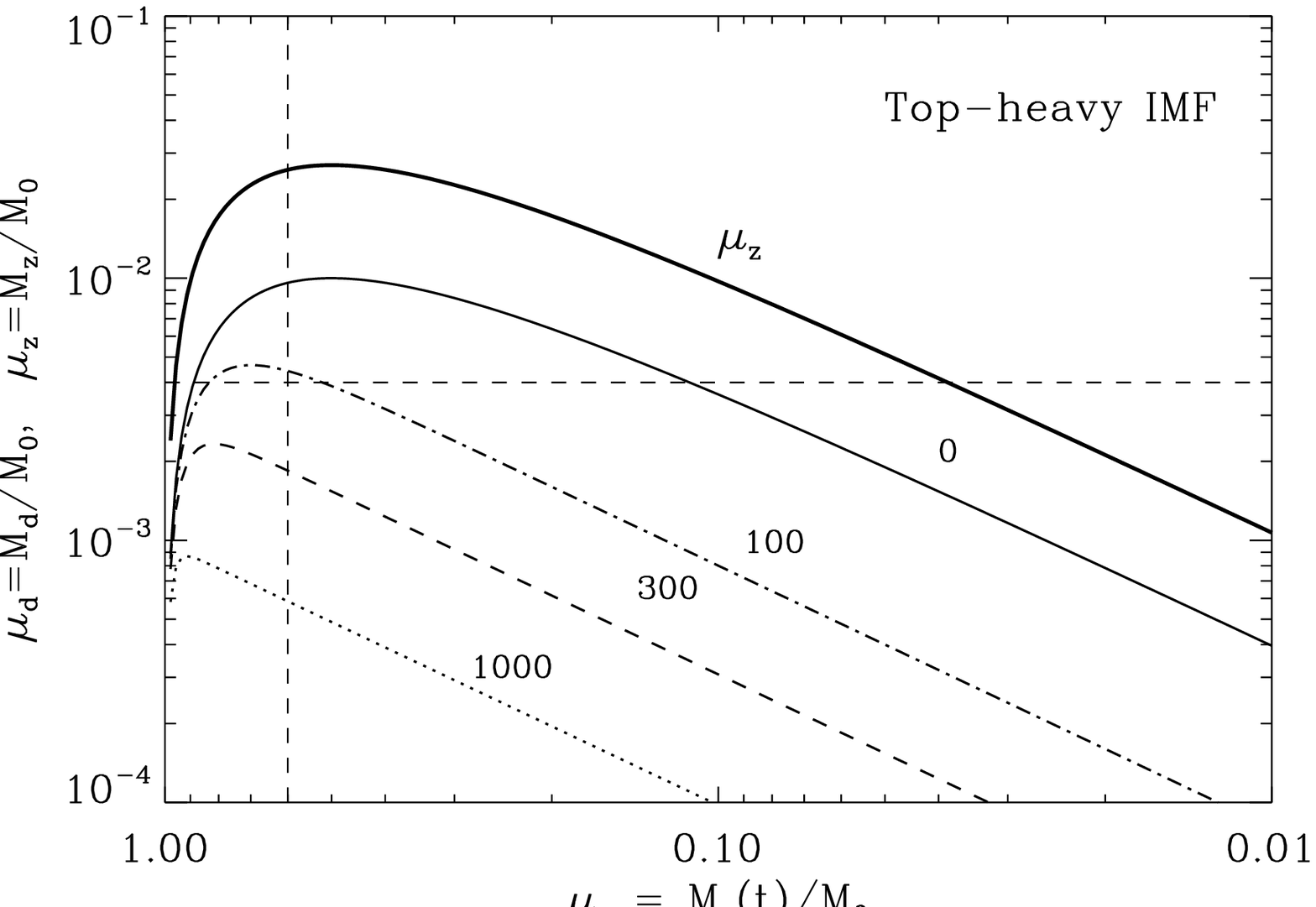} 
\end{tabular} 
\end{center}
 \caption{{\footnotesize The evolution of the mass of dust and metals, $M_d$ and $M_z$ respectively, both normalized to the initial gas mass, $M_0$,  as a function of  the fraction of the ISM gas, \mug, in a closed-box model for the chemical evolution.  Calculations are presented for a Salpeter IMF (left panel) and a top-heavy IMF (right panel).  Curves are labeled by \mism, the mass of ISM gas that is cleared of dust by a single SNR in units of \msun. A value of \mism\ = 0 corresponds to no grain destruction. The horizontal dashed line corresponds to the $M_d/M_0$ value of 0.004 adopted for \jay, and the vertical line to the value of \mug\ at the epoch of 400~Myr  (see Table \ref{adopt}).}}
   \label{mdmz_mu}
\end{figure}

Figure~\ref{mdmz_mu} depicts the evolution of the dust mass, $M_d(\muge)$ and mass of metals, $M_z$, normalized to the initial mass, $M_0$, for various values of \mism, the mass of ISM gas that is cleared of dust by a single SNR as a function of \mug. Initially, \mug = 1, but decreases as the gas is converted into stars. Calculations are presented for two different stellar IMF: a Salpeter IMF (left panel), and  a top-heavy IMF (right panel). Initially $M_d = 0$ and rises as the ISM is enriched by SN-produced dust. However, eventually the gas and dust in the ISM are incorporated into stars, and the mass of interstellar dust decreases. 

  \begin{figure}[h]
  \begin{center}
  \begin{tabular}{cc}
\includegraphics[width=3.0in]{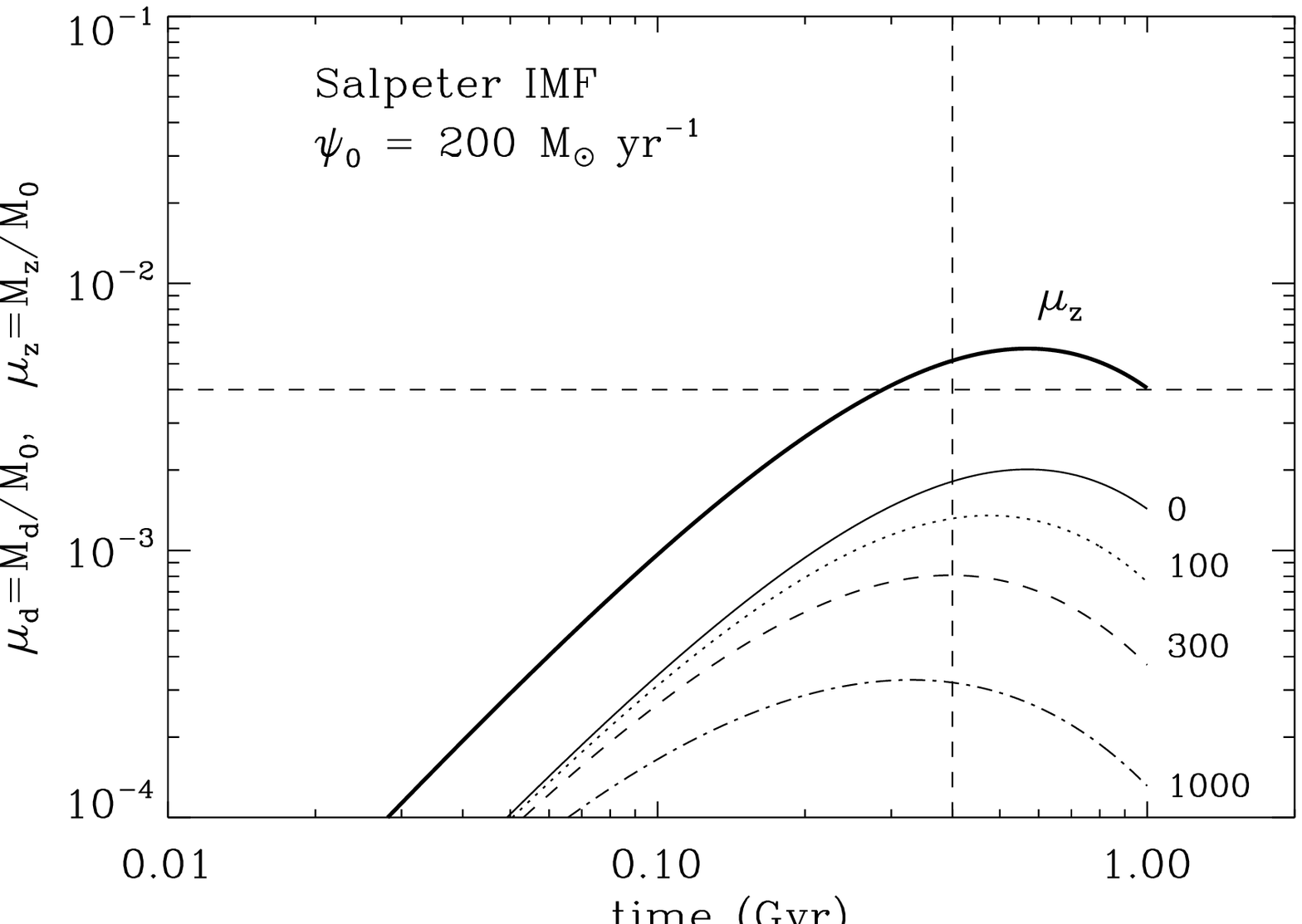}  &
\includegraphics[width=3.0in]{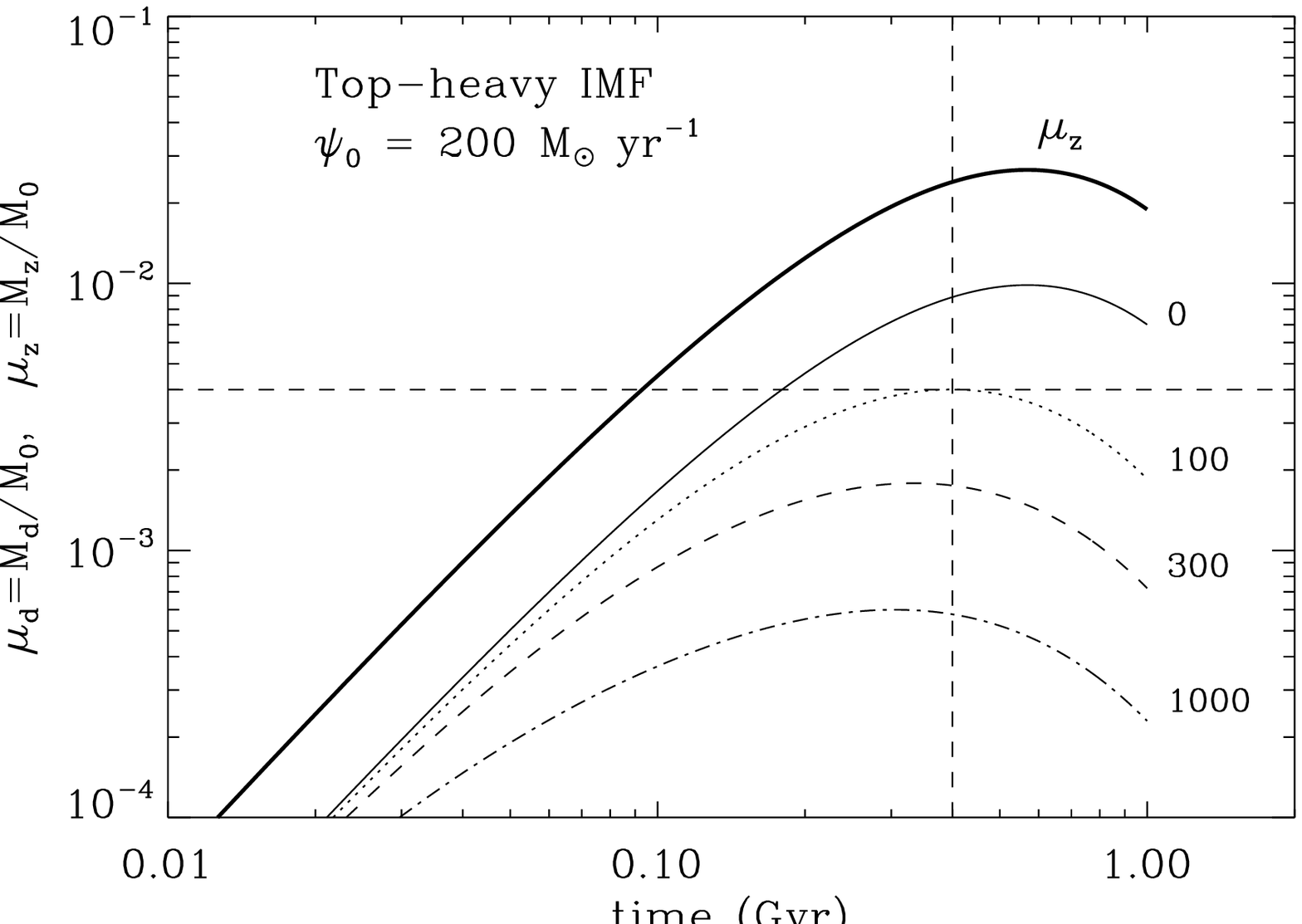} 
\end{tabular} 
\end{center}
 \caption{{\footnotesize The evolution of the mass of dust and metals, $M_d$ and $M_z$ respectively, both normalized to the gas mass, $M_0$,  at $t_0 = 400$~Myr as a function of time for the infall model.  Calculations are presented for a Salpeter IMF (left panel) and a top-heavy IMF (right panel). Curves are labeled by \mism\ in units of \msun.  The horizontal dashed line corresponds to the $M_d/M_0$ value of 0.004 adopted for \jay, and the vertical line corresponds to the adopted galaxy's age of 400~Myr  (see Table \ref{adopt}).}}
   \label{mdmz_mu_inf}
\end{figure}

Figure~\ref{mdmz_mu_inf} presents the same quantities for the infall model.  
Both figures show the maximum values of $\mu_d$ and $\mu_z$ attainable with each IMF. Larger values of $\mu_d$ and $\mu_z$ are obtained with a top-heavy IMF. The figures also show that without any grain destruction, the mass of dust is simply proportional to the mass of metals, but decreases more rapidly than the mass of metals when grain destruction is taken into account. 

\subsection{The Evolution of the Dust-to-Metals and Dust-to-Gas Mass Ratios}

\begin{figure}[htbp]
  \begin{center}
    \begin{tabular}{cc}
\includegraphics[width=3.0in]{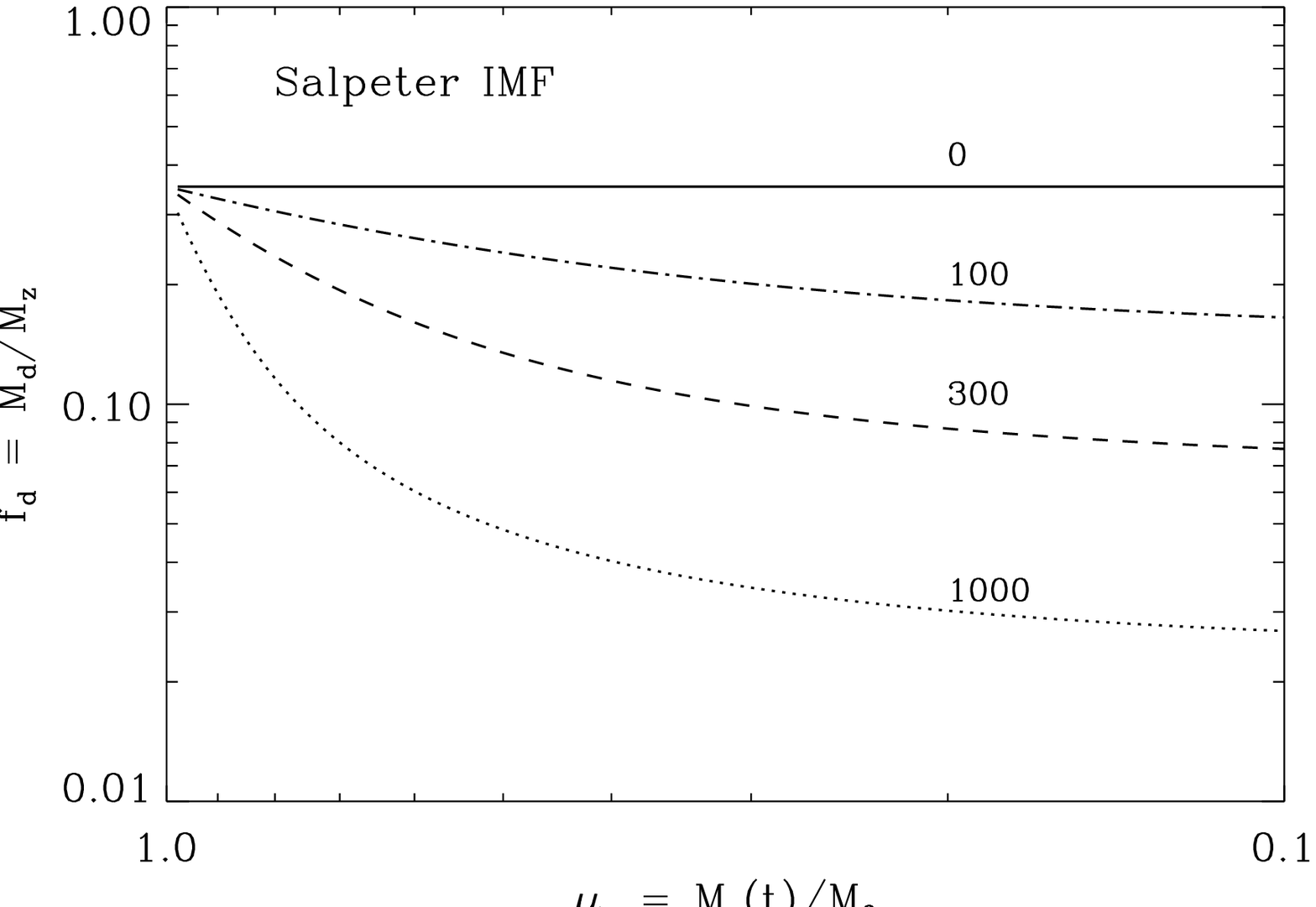}  &
\includegraphics[width=3.0in]{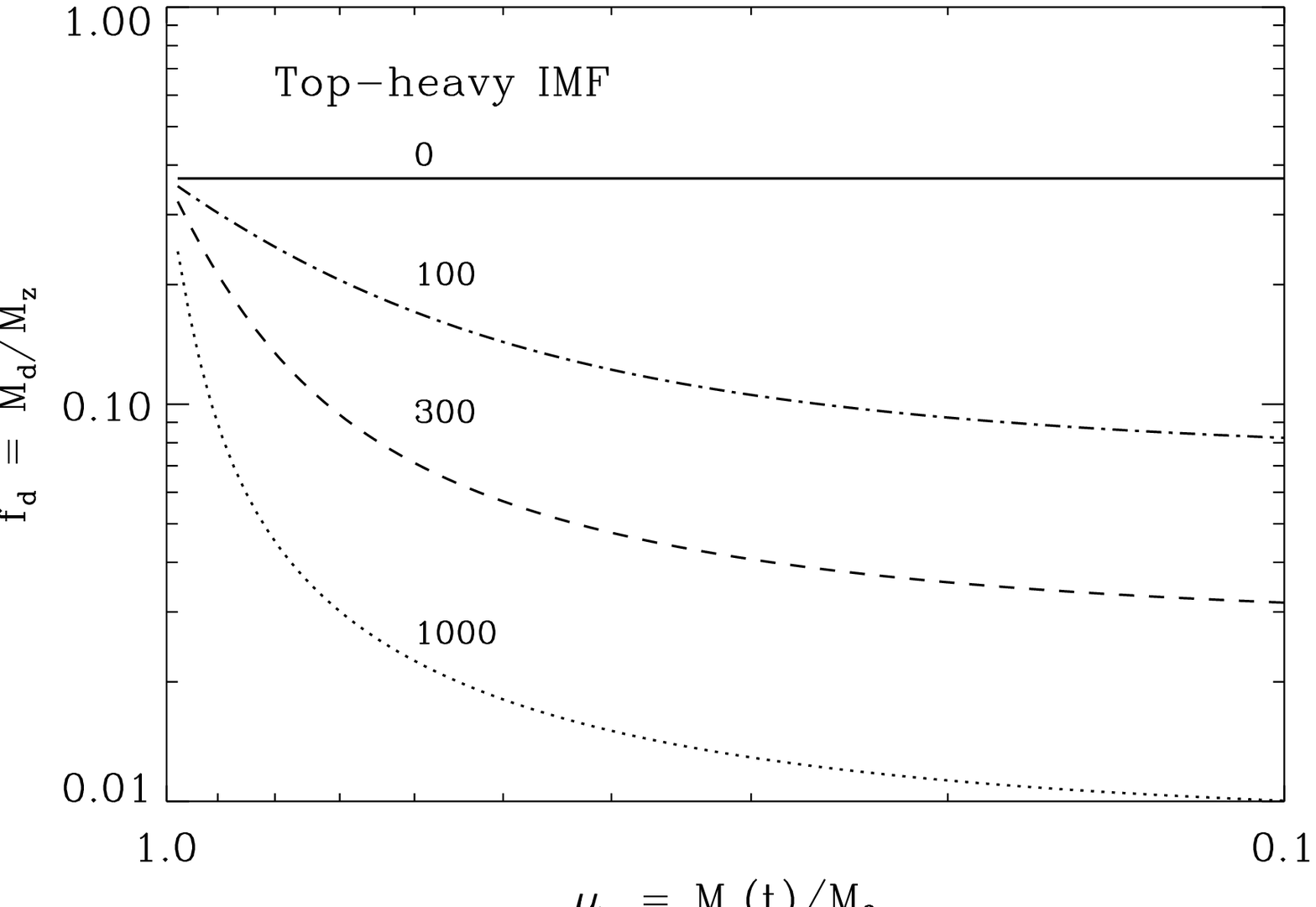} 
\end{tabular} 
\end{center}
 \caption{{\footnotesize The evolution of the mass fraction of  metals that is locked up in dust, $f_d$ given by eq. (20), is plotted versus \mug, the fraction of the ISM gas for the closed box model.  Curves are labeled by \mism\ in units of \msun. In the absence of grain destruction  the value of $f_d$ is equal to $\sim$ 0.35, the fraction of metals in the SN ejecta that condenses into dust. }}
    \label{fd_mu}
\end{figure}

The effect of grain destruction is to decrease the fraction of condensable elements in the solid phase of the ISM. This point is illustrated in Figure~\ref{fd_mu} which depicts the evolution of the mass fraction of metals locked up in dust, $f_d \equiv M_d/M_z$, versus \mug\ for the closed box model. When \mism\ = 0, $f_d$ is constant and equal to $\yde/\yze \approx 0.35$, the fraction of the metals in the SN ejecta that condensed and formed dust. As the figure illustrates, this fraction deceases with \mug\ as the grain destruction efficiency increases. Similar quantitative results can be obtained for the infall model.

\begin{figure}[h]
  \begin{center}
      \begin{tabular}{cc}
  \includegraphics[width=3.0in]{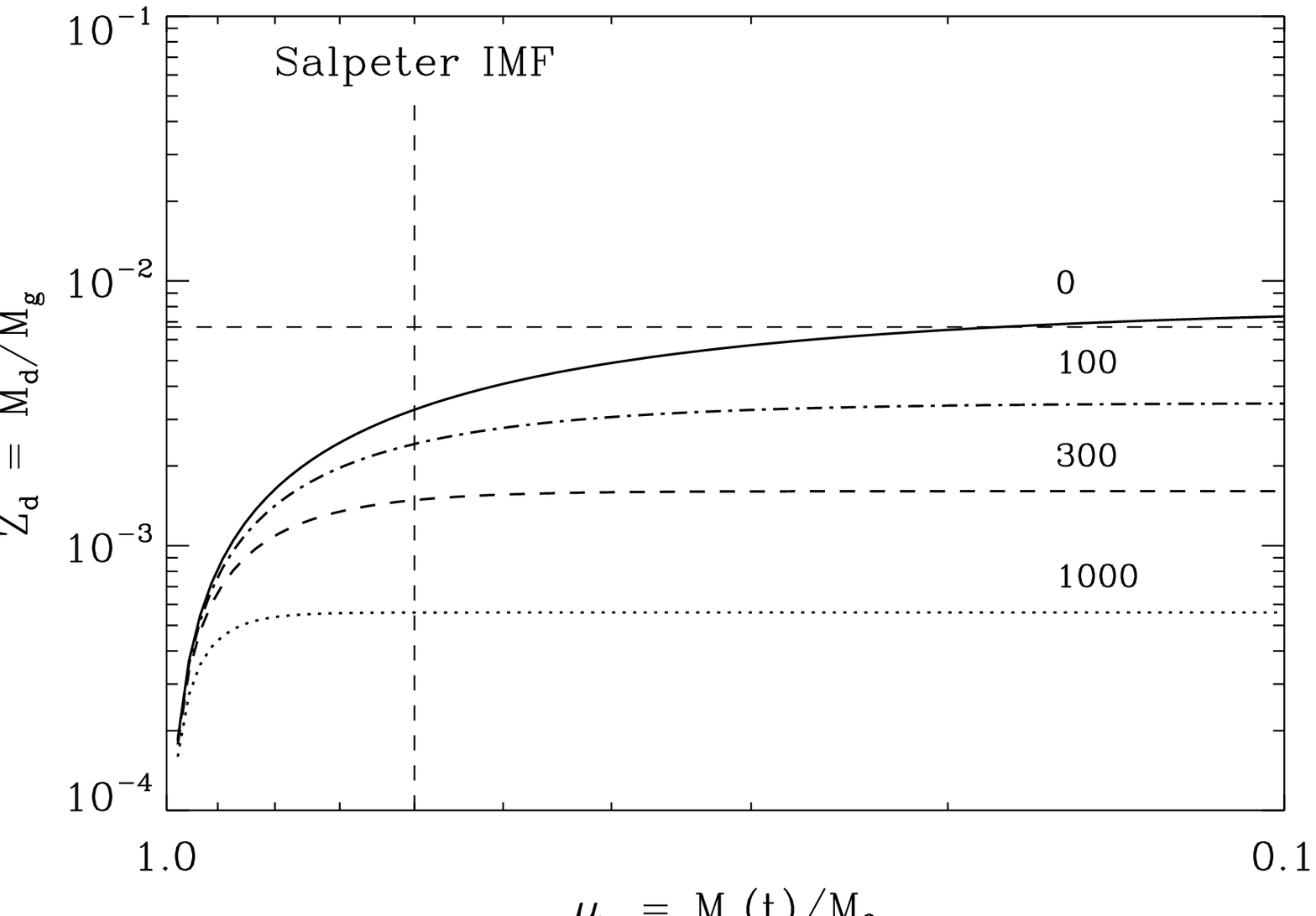} &
\includegraphics[width=3.0in]{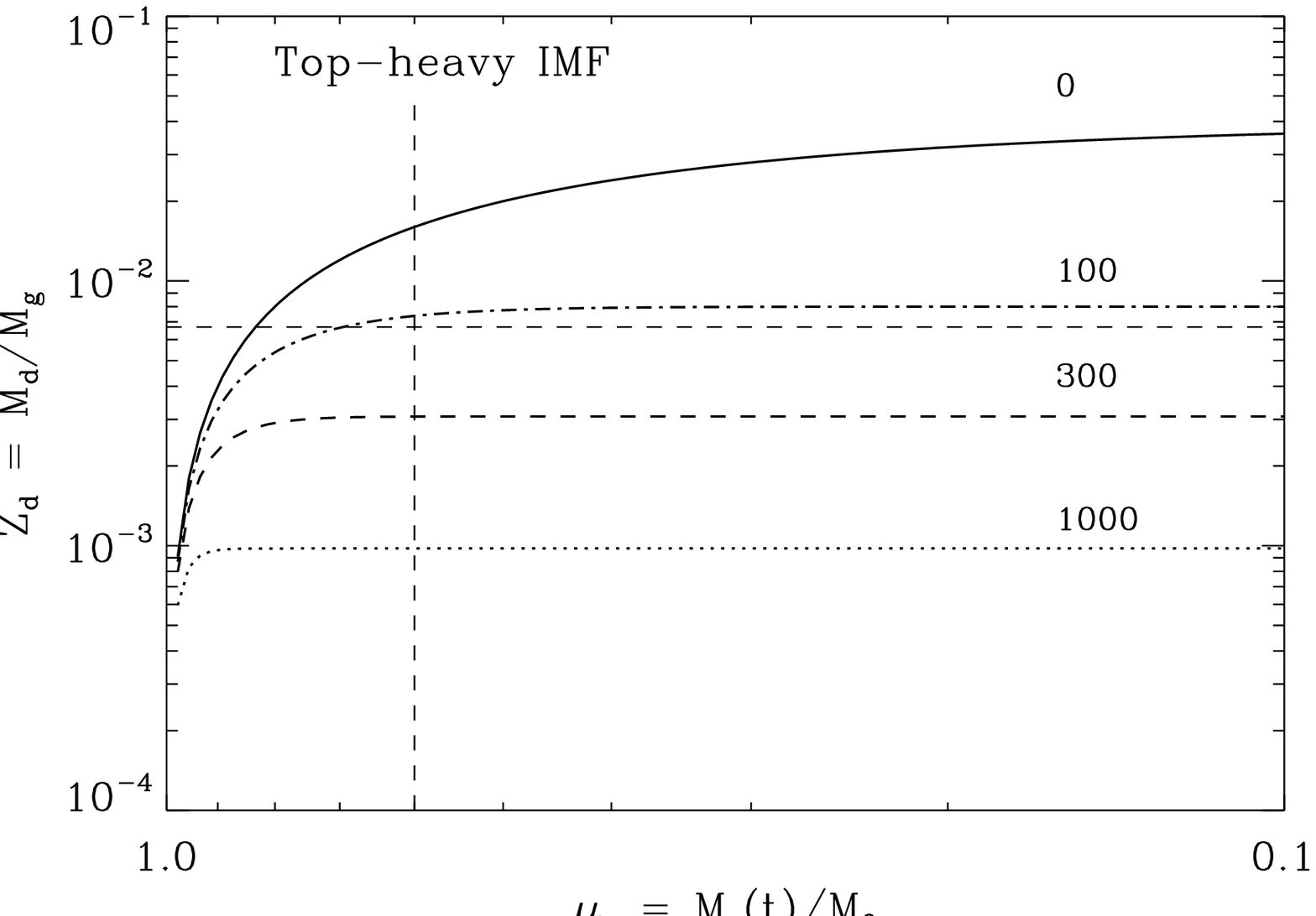} 
\end{tabular}
\end{center}
 \caption{{\footnotesize The evolution of the dust-to-gas mass ratio, $Z_d$, and the dust-to-metals mass ratio, $f_d$ as a function of \mug. Curves are labeled by \mism\ in units of \msun. The dashed horizontal and vertical lines represent, respectively, the adopted gas-to-dust mass ratio and gas mass fraction of \jay\ at $t=400$~Myr. The figure shows that when grain destruction is important, \mism $\gtrsim$ 100~\msun, a top-heavy IMF is required to produce the observed amount of dust at $\mu_g = 0.6$. }}
 \label{zd_mu}
\end{figure}

Figure~\ref{zd_mu} shows that when grain destruction is ignored, i.e. \mism\ = 0, the dust-to-gas mass ratio, $Z_d$, continues to rise since both the gas and dust are incorporated into stars, but the ISM is continuously enriched by dust formed in SN ejecta. When grain destruction is taken into account, $Z_d$ reaches a steady-state at values of \mug\ which become increasingly smaller as the grain destruction efficiency, which is related to the value of \mism, increases.

\subsection{The SN Dust Yields Needed to Produce an Observed Dust-to-Gas Mass Ratio }

\begin{figure}[htbp]
  \begin{center}
\includegraphics[width=5.0in]{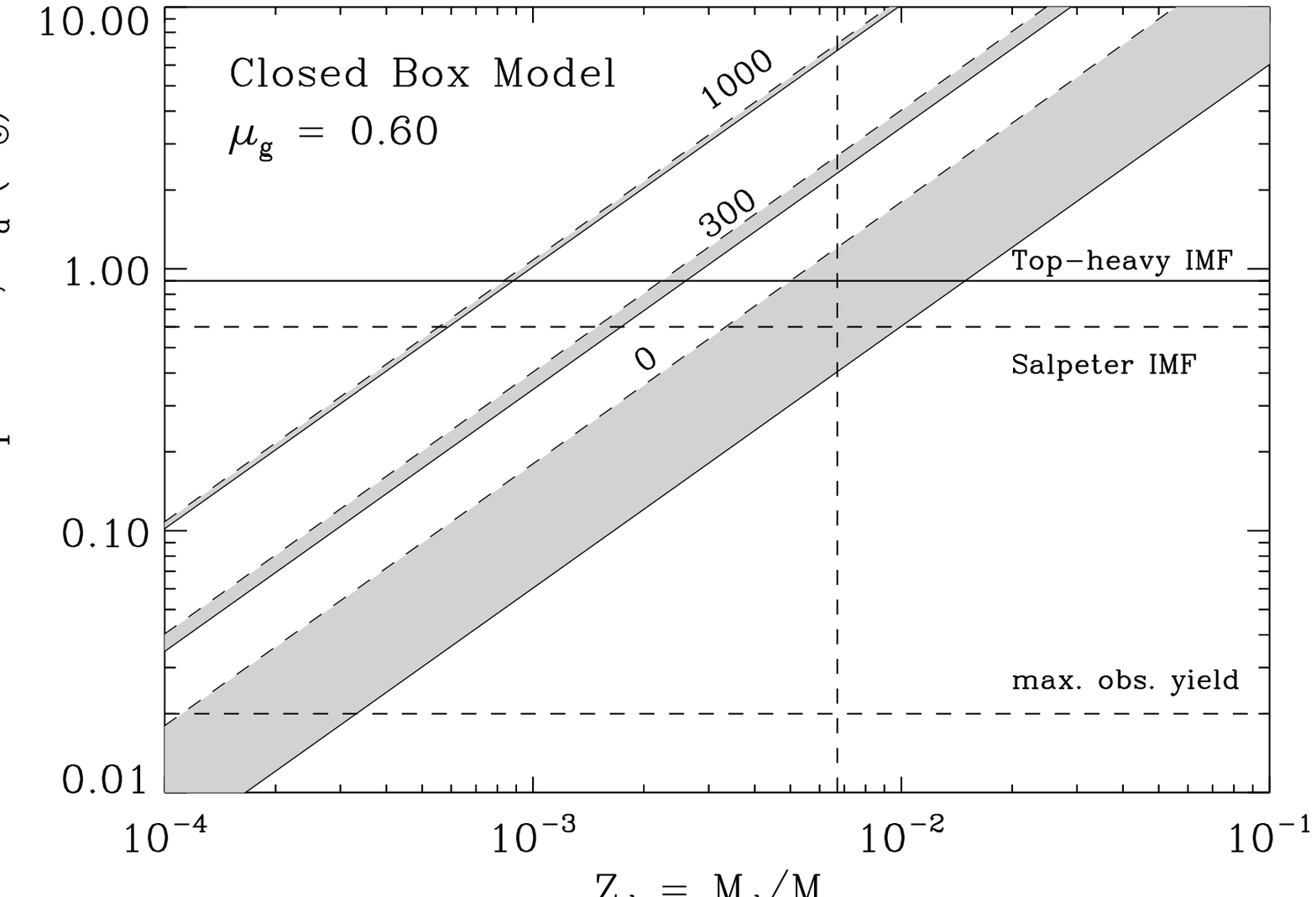} 
\end{center}
 \caption{{\footnotesize The IMF-averaged yield of dust by type~II supernova, \yd, that is required to account for a given dust-to-gas mass ratio $Z_d$, is presented for different values of \mism\ given in units of \msun. The horizontal dashed line near the bottom of the figure corresponds a value of $Y_d = 0.02$~\msun, the highest inferred yield of dust in a supernova to date (Sugerman et al. 2006). The vertical dotted line represents the value of $Z_d$ at $\muge =0.60$. Curves are labeled by \mism\ given  in units of \msun. Solid and dashed lines correspond to calculations done for a top-heavy and a Salpeter IMF, respectively. The top two dashed (solid) horizontal lines represent IMF-averaged theoretical dust yields for a Salpeter (top-heavy) IMF. The results are identical for both, the closed box and infall models.}}
    \label{ydzd}
\end{figure}

Figure~\ref{ydzd} shows how much dust an average SN {\it must} produce in order to give rise to a given dust-to-gas mass ratio, for various grain destruction efficiencies. The value of \yd\ was calculated when \mug\ reaches a value of 0.60, the adopted gas mass fraction of \jay\ at 400~Myr. The figure shows that, for example, to produce a value of $Z_d = 0.0067$ at \mug\ = 0.60, a SN must produce about 0.4~(1.2)~\msun\ of dust for a top-heavy (Salpeter) IMF, provided the dust is not destroyed in the ISM. Even with modest amount of grain destruction, \mism\ = 100~\msun, the required SN dust yield is dramatically increased to about $1-2$~\msun, depending on the IMF. 
The horizontal line in the figure corresponds to a value of \yd\ = 0.02~\msun, the largest amount of dust directly observed in the ejecta of a supernova \citep{sugerman06}. The figure shows that even without grain destruction, the largest observed yield can only give rise to a gas-to-dust mass ratio of $\sim 3\times10^{-4}$.

 
\section{APPLICATION TO J1148+5251}

The results of our chemical evolution model can be readily applied to any galaxy sufficiently young so that AGB stars are only minor contributors to the dust abundance in the ISM. Here we concentrate on the quasar \jay. 
 \subsection{Observational Properties}
 Table~\ref{j11_obs} summarizes the observed properties of  \jay. At redshift $z=6.4$ the age of the universe is 890~Myr for a $\Lambda$CDM universe with $\Omega_m = 0.27$, $\Omega_{\Lambda} = 0.73$ and a Hubble constant $H_0 = 70$~km~s$^{-1}$~Mpc$^{-1}$. 
  
\begin{figure}[htbp]
  \begin{center}
\includegraphics[width=4.0in]{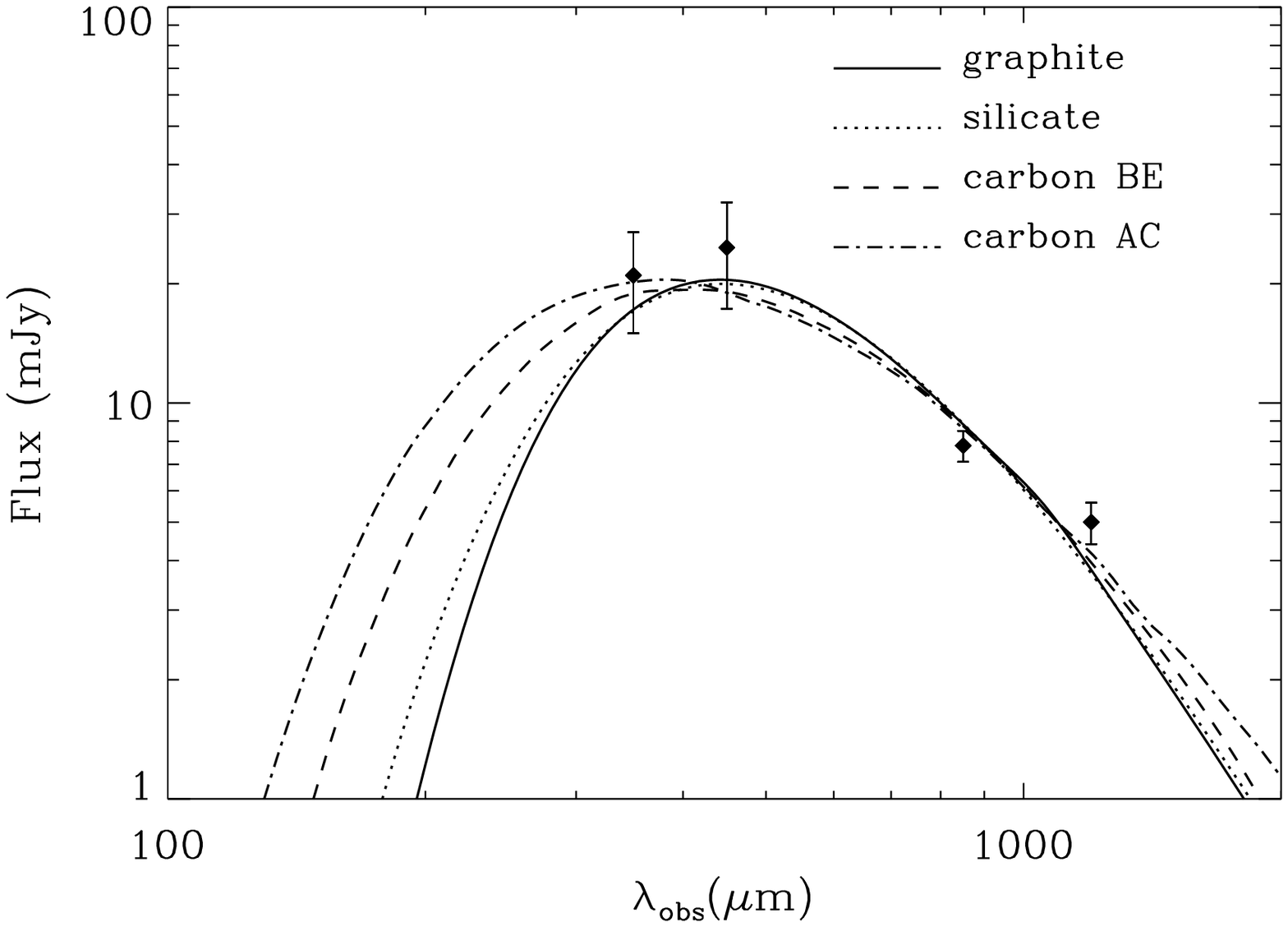}  
\end{center}
 \caption{{\footnotesize Spectral fits of several possible dust compositions to the observed far-IR and submillimeter observations of \jay. References to the observations are listed in Table~\ref{j11_obs}. Temperatures, masses, and luminosities of the different dust types are listed in Table~\ref{dustprop}. }}
    \label{dustspec}
\end{figure}

 Figure~\ref{dustspec} depicts the observed far-IR and submillimeter fluxes at the observed wavelengths. The different curves are spectral fits to these fluxes for different dust compositions. The optical properties for the silicate and graphite grains were taken from \cite{draine84} and for the carbon grains from Rouleau \& Martin (1991).  Dust masses vary from $\sim 10^8$ to $5\times 10^8$~\msun, depending on dust composition. Table~\ref{dustprop} summarizes the derived properties and IR luminosities for the different dust compositions.
 
For comparison with model calculations we assumed that the onset of star formation in this galaxy occurred at $z = 10$, when the universe was 490~Myr old, giving an age of 400~Myr for this object at $z=6.4$. Based on the CO observations (see Table \ref{j11_obs}) we adopted a molecular gas mass of $1.5\e{10}$~\msun, and and an equal mass of atomic gas, for a total gas mass of $3\e{10}$~\msun. From the far-IR emission we adopted an average dust mass of $M_d =  2\times 10^8$~\msun. From estimates of the dynamical mass, we adopted a total galactic mass of $M_0 = 5\times 10^{10}$~\msun\ at time $t\equiv t_0 = 400$~Myr. This mass corresponds to the initial mass of the galaxy in the closed box model, and to the total mass accreted until $t_0$ in the infall model. With these values, we get a gas mass fraction of $\muge = 0.60$, and a dust-to-gas mass ratio of $Z_d = 0.0067$ at $t = 400$~Myr. Finally, we adopted a far-IR luminosity of  $2\times 10^{13}$~\lsun. 
 Table~\ref{adopt} summarizes the various derived and adopted properties of \jay. 

 \subsection{The Dust Mass and the Required Dust Yield in Core-Collapse SN}
Figures \ref{mdmz_mu} and \ref{mdmz_mu_inf} give an upper limit on the mass of dust that can be produced in 400~Myr by supernovae when grain destruction is ignored, that is, when \mism\ = 0. The maximum dust mass is $\sim 10^8$~\msun\ ($\mu_d \approx 2\times10^{-3}$) for a Salpeter IMF, and $\sim 5\times10^8$~\msun\ ($\mu_d \approx 10^{-2}$) for a top-heavy IMF. The observed dust mass of  $2\times 10^8$~\msun\ requires therefore a top-heavy IMF, which can produce this mass of dust even with a grain destruction efficiency corresponding to a value of \mism\ = 100~\msun. This value of \mism\ corresponds to a grain lifetime of:
\begin{equation}
\tau_d = {M_g \over \misme\ R_{\rm SN}} = {3\times10^{10} \over 100\times 7.4} = 27~{\rm Myr}
\end{equation}
where the SN rate was calculated for a SFR of $10^3$~\myr and a Salpeter IMF for which \msn\ = 147~\msun.

This result is also illustrated in figure~\ref{ydzd} that shows that the yield of dust per SN required to produce a dust-to-gas mass ratio of 0.0067 with \mism\ = 100~\msun\ is about 1~\msun\ per SN. This is about equal to the total mass of condensable elements produced in a SN with a 25~\msun\ progenitor star (Woosley \& Weaver 1995; Nomoto et al. 2006). Such average dust yield is theoretically only attainable with a top-heavy IMF, provided that the dust condensation efficiency in the SN ejecta is about 100\%. Observationally, this yield is significantly higher than the $\sim 0.02$~\msun\ of dust in SNII~ 2003gd in the
galaxy NGC~628, inferred from the analysis of the IR emission and the internal extinction in the SN ejecta \citep{sugerman06}. If this low yield is typical, then SNe cannot be the dominant source of dust in these young galaxies. Alternative mechanisms that can produce the observed dust mass, such as accretion in molecular clouds, or formation around the AGN \citep{elvis02}, need then to be included in the dust evolution model.   

 \subsection{The Dependence of the Required SN Dust Production Yield on the Total Mass of the Galaxy}
The dust yield required to produce a given dust-to-gas mass ratio at time $t_0$ depends on the total mass of the galaxy, which we took to be the dynamical mass of \jay. Since this quantity is uncertain, we explore the dependence of the dust mass produced in the model on the adopted total mass of the system. For the closed box model, the relation between these two quantities, \yd\ and $M_0$,  is given by eq.~(\ref{md_mu}).

\begin{figure}[htbp]
  \begin{center}
\includegraphics[width=5.0in]{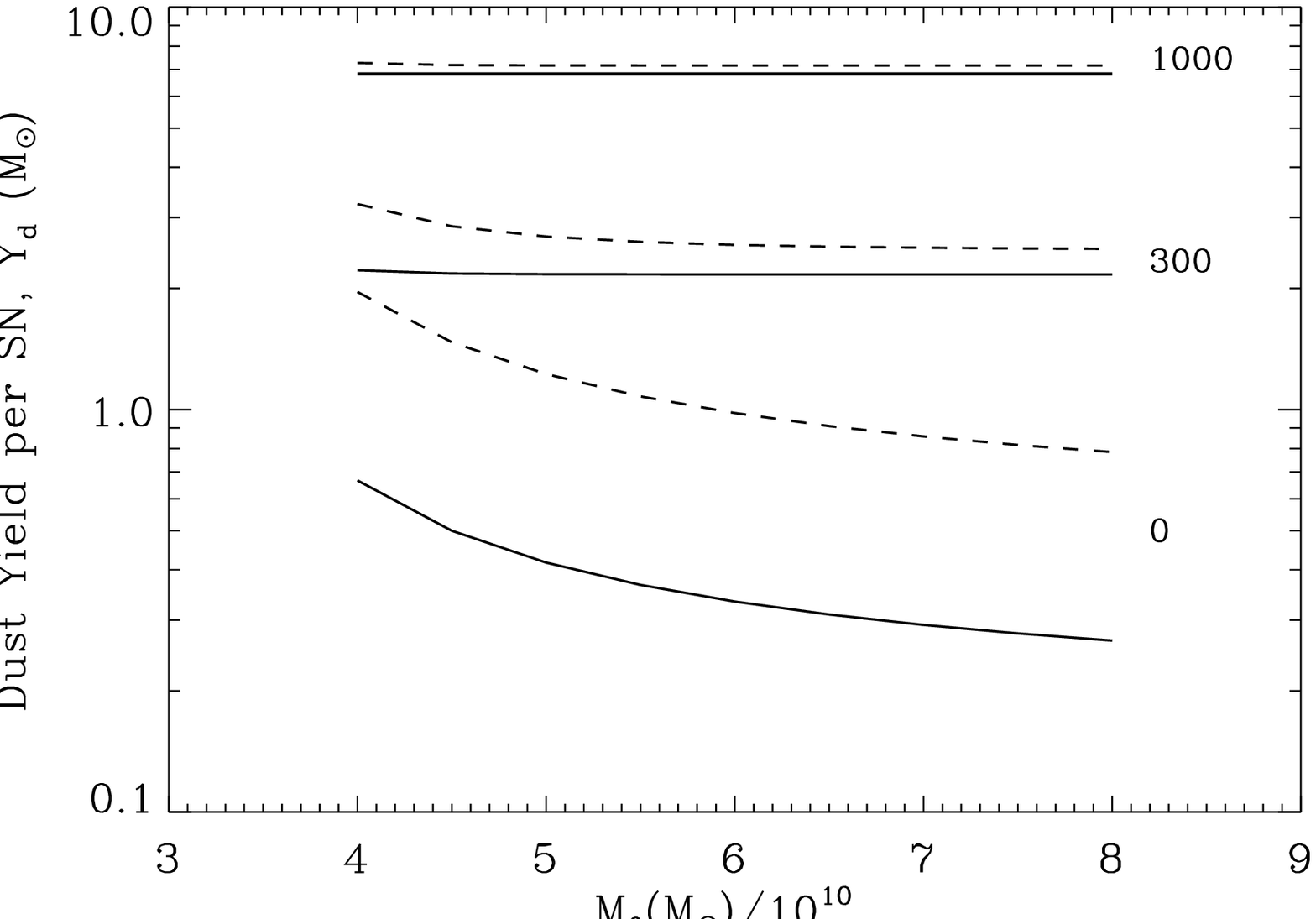} 
\end{center}
 \caption{{\footnotesize The dependence of the dust yield in an average Type~II supernova required to produce the observed dust mass of $2\times10^8$~\msun, on $M_0$, the total mass of the system (see eq.~\ref{md_eq}). The calculations were performed for a gas mass of $M_g = 3\times10^{10}$~\msun\ and a dust mass of $M_d = 2\times10^8$~\msun. Curves are labeled by \mism\ given  in units of \msun. Solid and dashed lines correspond to calculations done for a top-heavy and a Salpeter IMF, respectively. When grain destruction is taken into account, te required dust yield becomes independent of the total mass of the galaxy. }}
    \label{ydm0}
\end{figure}

The total mass of the galaxy must exceed its gas mass, which we took to be $3\times10^{10}$~\msun, so we varied $M_0$ from 4 to $8\times 10^{10}$~\msun, which spans the dispersion in the observed dynamical mass of the galaxy. Figure \ref{ydm0} depicts the results of our calculations for the two different IMFs and the three different values of \mism\ used in figure \ref{ydzd}. In the absence of grain destruction \yd\ decreases by a factor of $\sim 4-5$ over the entire mass range of $M_0$. However, when grain destruction is taken into account, \yd\ becomes increasingy independent of the total galactic mass, approaching  the asymptotic behavior for $\misme \gg R $~\mstar\ presented in eq.~(\ref{yd_asymp}). For example, for $\misme \approx 300$~\msun, and a dust-to-gas mass ratio of $Z_d \approx 0.003$ we get that $\yde \gtrsim 1$~\msun, significantly larger than any dust mass observed in any SN ejecta. 

 \subsection{The Star Formation Rate in \jay}

From the observed luminosity and gas mass we can derive the current SFR using the relations between the SFR and the far-IR luminosity and the total gas mass \citep{kennicutt98a,kennicutt98b}.
For a Salpeter IMF the SFR is related to the far-IR luminosity of the galaxy by:
\begin{equation}
\psi(M_{\odot}~yr^{-1})  =  1.7\times 10^{-10}\ L(L_{\odot}) \qquad {\rm Salpeter\ IMF} 
\end{equation}
which for a value of $L_{\rm IR} = 2\times 10^{13}$~\lsun, gives a SFR of $\sim 3400$~\myr. A high SFR of $\sim 3000$~\myr\ was also derived by \cite{maiolino05} from the luminosity in the $\left[{\rm C\ II}\right]$~158$\,$\mic\ line detected in this quasar. 

The SFR  can also be derived from the gas mass using the empirical relation between the SFR per unit area, $\Sigma_{\rm SFR}$, and  the gas mass surface density, $\Sigma_g$ in star forming galaxies (Kennicutt et al. 2006):
\begin{equation}
\Sigma_{\rm SFR}\ ({\rm M}_{\odot}\ {\rm yr}^{-1}\ {\rm kpc}^{-2}) = (4.5^{+1.0}_{-0.84})\times 10^{-5}\ \left({\Sigma_g \over  {\rm M}_{\odot}\  {\rm pc}^{-2}}\right)^{1.56\pm 0.04}  \qquad {\rm Salpeter\ IMF} 
\end{equation}
The CO maps of \jay\ suggest that it consists of two blobs of comparable size with a diameter of about 1~kpc. For an adopted gas mass of $3\times10^{10}$~\msun\ we get a total SFR $\psi \simeq 300$~\myr. 

The two different tracers give very different star formation rates. Both estimates are highly uncertain: the first assumes that the dust reradiates all the starburst's luminosity and ignores any possible contribution of an AGN to the heating of the dust. The second assumes that the CO observations traces the densest gas regions in the object, as inferred from HCN-CO correlations, and is therefore a good measure of the total gas mass in galaxies \citep{gao04}. However, HCO$^+$ observations which probe denser regions of molecular gas, have cast doubts on the reliability of HCN (and hence CO) as an unbiased tracer of dense molecular gas in ULIRGs \citep{gracia-carpio06}.
 
Another major source of uncertainty is the stellar IMF. For example, the SFR derived from the IR luminosity using a Salpeter IMF, will decrease from a value of $3400$~\myr, to a rate of about 380 \myr\ for a mass- or top-heavy  IMF. Because of all these uncertainties, the star formation rate of \jay\ cannot be uniquely determined. In the following, we discuss the dependence of its star formation history on the assumed current SFR.

 \subsection{The Star Formation History of \jay}
In the closed box model, the evolution of the dust, and metals could be presented as a function of the gas mass fraction \mug\ (Figures~\ref{mdmz_mu}--\ref{zd_mu}). The translation of the dependence of these quantities from \mug\ to time requires knowledge of the star formation history of \jay, which in turns depends on the assumed SFR and the initial gas mass of the object. For example, given an initial gas mass of $M_0 = 5\times 10^{10}$~\msun, a SFR of 3000~\myr\ will exhaust almost all the available gas in less than $\sim 100$~Myr. This suggests one or more of the following possibilities: (1) the onset of star formation occurred relatively shortly before the observations; (2) the galaxy started with an initially larger reservoir of mass;  (3) that rapid infall, comparable to the SFR, kept the reservoir of gas sufficiently high.

Some of these possibilities were presented more qualitatively in figure~\ref{mutime}, which depicts the evolution of the mass fraction of gas as a function of time for the closed box and infall models. For the closed box model (left panel), the calculations assume an initial gas mass of $5\times 10^{10}$~\msun, and the Kennicutt law for the relation between the SFR and the gas mass: $\psi(t) \propto M_g(t)^k$, with $k = 1.5$ [see eq.~(\ref{mg_eq})]. The solid line depicts the evolution of $\muge(t)$ for an initial SFR $\psi_0 = 150$~\myr, corresponding to a value of $\psi = 70$~\myr\ at $t = 400$~Myr, which reproduces the adopted value of $\muge = 0.60$ at that epoch. The additional curves depict the evolution of $\muge(t)$ for initial SFRs of $650$,   $2.2\times10^3$, and  $6.5\times10^3$ \myr, corresponding to values of $\psi(t_0) =$ 300, $1\times10^3$, and $3\times10^3$~\myr, at $t_0 = 400$~Myr. The corresponding gas mass fractions at that epoch are: 0.19, 0.036, and 0.0052, respectively.
The figure shows that a low observed SFR of only 70~\myr\ is required to fit the observations of \jay, given the initial conditions and assumptions summarized in Table~\ref{adopt}. A current SFR of 3000~\myr\ requires changes in the initial conditions and model assumptions. If the galaxy is indeed 400~Myr old, then the value of \mug\ is 0.0052, requiring the initial gas mass of \jay\ to be $5\times 10^{10}/0.0052 \simeq 1\times10^{13}$~\msun. A more plausible scenario is that \mug\ is 0.60, but that the age of the starburst is only about $10^7$~yr.

Similar conclusions are reached for the infall model. The right panel of Figure~\ref{mutime} depicts the evolution of the  gas mass, constrained to fit the adopted values of $M_0$ and $\mu_g(t_0)$ at time $t_0=400$~Myr. The fit requires the values of the product $\psi_0\, t_0$ to be between $\sim(5-11)\times10^{10}$~\msun.
If star formation has been an ongoing process over a period of 400~Myr, then the current SFR must be between $\sim$ 125 and 285~\myr. To accommodate a much larger SFR, say of 3000~\myr, requires the age of the starburst to be about $3\times10^7$~yr.

 \subsection{The Spectral Energy Distribution of \jay}
\begin{figure}[htbp]
  \centering
  \includegraphics[width=\textwidth]{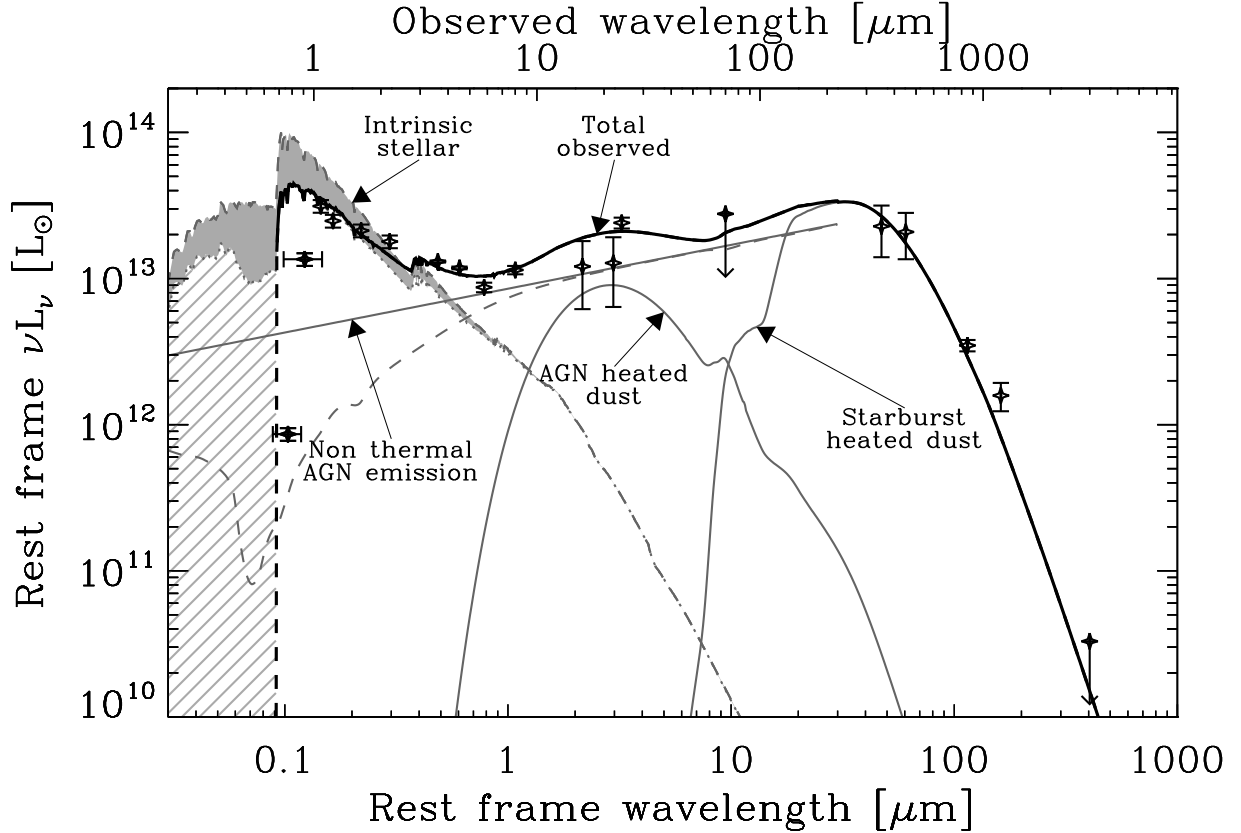}
  \caption{{\bf SED of \jay}.
           The observed fluxes from \jay\ are plotted as a function of the rest frame wavelength for a redshift of 6.4. Data and references are given  in Table \ref{tab:SED}. The galaxy's SED consists of four distinct emission components: the energy released by the starburst; the emission from the AGN; the starlight absorbed by the dust and reradiated at far-IR wavelengths; and the emission from the AGN that is absorbed by the dust and reradiated at mid-IR wavelengths. The gray area in the figure represents the stellar energy absorbed by the dust, and the hatched area the fraction of the ionizing stellar radiation that is absorbed by the gas. Details on the derivation of the emission components are given in \S3.5 of the text.}
\label{sedJ11}
\end{figure}

The SED of \jay\ offers very few clues regarding the relative starburst or AGN contribution to the thermal dust emission from this galaxy. Figure~\ref{sedJ11} depicts the galaxy's SED from UV to submillimeter wavelengths. The  UV and optical parts of the spectrum are most likely dominated by escaping starlight, and the far-IR by reradiated thermal emission from dust. The dashed-dotted grey line in the figure represents the intrinsic stellar radiation field synthesized with P\'EGASE \citep{fioc97} for a continuous star formation rate of age $t=400$~Myr, with a top-heavy IMF, and a SFR of 2500~\myr. The total intrinsic stellar luminosity is $1.0 \times 10^{14}$~\lsun. 
Part of this stellar energy is absorbed by dust and reradiated at IR wavlengths. We used  a simple screen model with a Galactic extinction law \citep{zubko04} to calculate the spectrum of the escaping stellar radiation, depicted by the dotted line in the figure. The magnitude of the extinction was chosen so that the total energy absorbed by the dust, shown as a grey shaded area in the figure, is equal to the total reradiated far-IR emission. The total luminosity radiated by the starburst-heated dust is $4.6\times10^{13}$~\lsun. The composition of this dust was taken to consists of a mixture of silicate and graphite dust with mass fractions of $2/3$ and $1/3$, respectively, and a $T^{-6}$ distribution of dust temperatures ranging from 40 to 150~K. 

The thick solid line represents the sum of all emission components, and represents the best $\chi^2$ fit of select model parameters (the intensity of the starburst, and the slope of the power law describing the AGN spectrum) to the observations. The model described above is only a plausible one, and definitely not unique. The possible existence of many distinct emission components, and the uncertainty in the IMF  illustrate the difficulty in determining the star formation rate from the galaxy's SED.

 The origin of the near- to mid-IR (NMIR) emission is more uncertain. The rest-frame $\sim 0.5 - 1$~\mic\ fluxes are in excess of the stellar emission that can be produced by a young starburst. It also cannot be produced by dust, since it requires the grains to radiate at temperatures above their sublimation point of $\sim 1500$~K. We therefore fit the near- to mid-IR emission with two components: an AGN represented by a $\nu^{-1.3}$ power law, and a hot dust component represented by a 1:9 mix (by weight) of silicate and graphite grains with a $T^{-6}$ distribution of dust temperatures ranging from 150 to 1500~K. The relatively low silicate-to-graphite mass ratio was chosen to avoid the production of a mid-IR excess due to the 9.7 \mic\ silicate emission feature. The power law is in good agreement with the average value found in the sample of
           ISO Palomar-green QSOs studied by \citet{haas03}.  
 The total intrinsic luminosity of the AGN is about $7\times 10^{13}$~\lsun, and the luminosity radiated by the AGN-heated dust is $1.3\times 10^{13}$~\lsun.
 
If the black hole (BH) radiates at the Eddington luminosity:
\begin{equation}
L_{edd}(L_{\odot}) \approx 3\times 10^{4}\ \left({M_{\scriptsize BH}\over M_{\odot}}\right)
\end{equation}
then the BH mass required to produce the AGN luminosity is $\sim 2\times10^9$~\msun, comparable to the mass estimate derived by \cite{willott03} from the width of the Mg~II line.
Mechanisms for the formation of seed black holes that enable their rapid growth to masses in excess of $\sim 10^9$~\msun\ have been discussed by \cite{lodato06}.

\section{SUMMARY AND DISCUSSION}
The early universe is a unique environment for studying the role of massive stars in the formation and destruction of dust. 
In this paper we developed analytical models describing the evolution of the gas, dust, and metallicity in high redshift galaxies. The equations describing their chemical evolution can be greatly simplified by using the instantaneous recycling approximation, and by neglecting the delayed contribution of low mass stars to the metal and dust abundance of the ISM. Neglecting any accretion of metals onto dust in the interstellar medium, the evolution of the dust is then completely driven by the condensation of refractory elements in the ejecta of Type~II supernovae, and the destruction by SN blast waves in the interstellar medium. 
The solutions for the evolution of the mass of gas, dust, and metals are presented in \S2 for closed box and infall models for the chemical evolution of the galaxy, and for different functional forms for the stellar initial mass function.
The results of our paper can be briefly summarized as follows:

\begin{enumerate}
\item The maximum attainable dust-to-metal mass ratio in any system is equal to the IMF-averaged mass fraction of metals that are refractory and able to condense onto grains in SN ejecta, which is about 0.35 (see Table \ref{dustyield}).
\item This mass fraction is significantly reduced when grain destruction is taken into account (Figure~\ref{fd_mu}). An observed dust-to-metals mass fraction $\gtrsim 0.4$ will therefore imply  that  accretion of ices onto interstellar grains in the ISM may be important in determining the dust mass in the galaxy.
\item Grain destruction plays an important role in the evolution of dust. However, its efficiency depends on the morphology of the ISM and is therefore highly uncertain (\S2.6). We therefore present all our results for different values of \mism, the effective mass of ISM gas that is completely cleared of dust by a single SNR.
\item In \S3 we present the general results of our models, describing the evolution of the gas, the dust, and the metals for both, the closed box and infall models.
\item We applied our general results to \jay, a dusty, hyperluminous quasar at redshift $z = 6.4$.
The observed and adopted quantities of \jay\ are summarized in Tables \ref{j11_obs}, \ref{dustprop}, and \ref{adopt}.  
\item The formation of about $2\times 10^8$~\msun\ of dust in this galaxy requires an average SN to produce about 1~\msun\ of dust (Fig. \ref{ydzd}). Theoretically, such large amount of dust can be produced if stars are formed with a top-heavy IMF, and with a moderate amount of grain destruction (\mism $\approx 100$~\msun). A Salpeter IMF fails to produce this amount of dust even in the absence of any grain destruction. Observationally, the required dust yield is in excess of the largest amount of dust ($\sim 0.02$~\msun) observed so far to have formed in a SN. This suggests that accretion in the ISM may play an important role in the growth of dust mass.
\item Figure~\ref{sedJ11} depicts the galaxy's spectral energy distribution from UV to far-IR wavelengths. The SED includes emission from the starburst, the AGN, and hot and cold dust components,  radiating at mid- and far-IR wavelengths, respectively.
\item Uncertainties in the fraction of the infrared luminosity that is powered by the starburst and in the stellar IMF prevent any accurate determination of the current star formation rate in the galaxy, or the unique determination of its star formation history (see \S3.4 and Fig. \ref{mutime}). 
\item  Simple decomposition of the galaxy's SED into its emission components suggest that the intrinsic starburst luminosity is about $1\times 10^{14}$~\lsun, $4.6\times 10^{13}$~\lsun\ of which is absorbed and reradiated by dust at far-IR wavelengths.
\item The $\sim 3$~\mic\ IR emission from the galaxy can neither be produced by starlight nor hot dust. It therefore must be emission from the AGN, and we estimate the AGN luminosity to be about $7\times 10^{13}$~\lsun, $1.3\times 10^{13}$~\lsun\ of which is assumed to be absorbed and reradiated by dust at mid-IR wavelengths.  
\item The AGN luminosity requires the formation of a black hole of a mass $\gtrsim 2\times 10^9$~\msun\ at this redshift.
\end{enumerate}

{\acknowledgements}
We thank Brad Gibson and the anonymous referee for comments that have contributed to the improvement of the manuscript. E.D. acknowledges the support of NASA's LTSA03-0000-065. 
The work of F.G. was supported by Research Associateship awards from the National
Research Council (NRC) and from the Oak Ridge Associated Universities (ORAU) at NASA Goddard Space Flight Center.

\newpage

\bibliography{/Users/edwek/science/00-Bib_Desk/Astro_BIB.bib}

\newpage

\begin{deluxetable}{llllllllll}
\tablewidth{0pt}
\tabletypesize{\footnotesize}
\tablecaption{Values of IMF-averaged Quantities$^1$}
\tablehead{
\colhead {IMF } &
\colhead {} &
\colhead {$\alpha$  } &
\colhead {$m_{low}$} &
\colhead { $m_{up}$ } &
\colhead { \mav } &
\colhead { \mstar } &
\colhead{\fsn} &
\colhead{\yz} &
\colhead{\yd} 
}
\startdata
Salpeter		& & 2.35 &  0.1 &   100 & 0.35 &  147 & 0.0024 & 1.4(1.7) & 0.5 (0.6) \\ 
mass-heavy  	&  & 2.35  &  1.0 &   100 &  3.1  &    58 & 0.054 & 1.4 (1.7) & 0.5 (0.6)\\
top-heavy   	&  & 1.50  &  0.1 &  100 &  3.2   &    50 & 0.064 & 2.2 (2.7) & 0.7 (0.9)\\
\enddata
\tablenotetext{1}{See \S2.1 for the definition of all quantities. Masses and yields are in units of \msun. Metallicity and dust yields were calculated for metallicities of 0.01\zsun, and \zsun\ (in parenthesis).  }
\label{imf}
\end{deluxetable}

\begin{deluxetable}{llllllll}
\tablewidth{0pt}
\tabletypesize{\footnotesize}
\tablecaption{Maximum Dust Yield in Massive Stars}
\tablehead{
\colhead {Mass} &
\colhead{$Z/Z_{\odot}$} &
\colhead {Metals} &
\colhead {Silicates} &
\colhead {Carbon} &
\colhead {Ca-Ti-Al} &
\colhead {Dust } &
\colhead {$Y_d/Y_z$} 
}
\startdata
12. &  0.01 & 0.4701 &  0.2848 &  0.0901 &  0.0043 &  0.3792 &  0.8066 \\
	&  1.0 & 0.6712 &  0.2697 &  0.0815 &  0.0235 &  0.3747 &  0.5583 \\
13. &  0.01 & 0.6413 &  0.3670 &  0.1090 &  0.0077 &  0.4837 &  0.7542 \\
	&  1.0 & 0.7869 &  0.3691 &  0.1151 &  0.0090 &  0.4932 &  0.6268 \\
15. &  0.01 & 0.9487 &  0.4164 &  0.1490 &  0.0110 &  0.5764 &  0.6075 \\
	&  1.0 & 1.4340 &  0.4759 &  0.1623 &  0.0248 &  0.6630 &  0.4623 \\
18. &  0.01 & 1.6982 &  0.4414 &  0.1940 &  0.0162 &  0.6516 &  0.3837 \\
	&  1.0 & 2.1711 &  0.5337 &  0.2493 &  0.0288 &  0.8118 &  0.3739 \\
20. &  0.01 & 2.4372 &  0.5689 &  0.2080 &  0.0282 &  0.8052 &  0.3304 \\
	&  1.0 & 3.0787 &  0.7910 &  0.2143 &  0.0287 &  1.0341 &  0.3359 \\
22. &  0.01 & 2.9129 &  0.8025 &  0.2480 &  0.0333 &  1.0839 &  0.3721 \\
	&  1.0 & 3.7869 &  1.1356 &  0.2424 &  0.0414 &  1.4193 &  0.3748 \\
25. &  0.01 & 4.0769 &  1.0311 &  0.2790 &  0.0333 &  1.3435 &  0.3295 \\
	& 1.0 &  5.0297 &  1.1484 &  0.3234 &  0.0704 &  1.5422 &  0.3066 \\
30. &  0.01 & 6.0489 &  1.3282 &  0.3150 &  0.0439 &  1.6871 &  0.2789 \\
	& 1.0 &  7.2639 &  2.0342 &  0.2916 &  0.1159 &  2.4417 &  0.3361 \\
35. &  0.01 & 8.3389 &  1.9632 &  0.3520 &  0.0494 &  2.3647 &  0.2836 \\
	 &  1.0 & 9.7242 &  2.5560 &  0.3216 &  0.1590 &  3.0366 &  0.3123 \\
40. & 0.01 & 10.5787 &  2.3926 &  0.3890 &  0.0722 &  2.8538 &  0.2698 \\
	 & 1.0 & 11.8819 &  3.0309 &  0.3696 &  0.1919 &  3.5924 &  0.3023
\enddata
\tablenotetext{1}{Based on \cite{woosley95} yields of massive stars. }
\label{dustyield}
\end{deluxetable}

\begin{deluxetable}{lll}
\tablewidth{0pt}
\tabletypesize{\footnotesize}
\tablecaption{Observed Properties of the QSO J1148+5152}
\tablehead{
\colhead { Observed quantity} &
\colhead { Value} &
\colhead{Reference} 
}
\startdata
 & & \\
      R.A. (J2000)     & $11^h48^m16\fs6$             & \citep{fan03}       \\
      Dec. (J2000)     & $+52\degr51\arcmin50\arcsec$ & \citep{fan03}       \\
      $\theta$		& $0.2\arcsec$ & \citep{walter04} \\
   & & \\
    \hline
 & & \\
      $z$(\lya)   & $6.37\pm0.03$                & \citep{white03}     \\
      $z$([Mg~II]) & $6.403\pm0.005$              & \citep{iwamuro04}   \\
      $z$(CO)     & $6.419\pm0.001$              & \citep{bertoldi03b}  \\
       & & \\
    \hline
     & & \\
     $M$(CO(3--2)) 
                       & $\sim1.6\times10^{10}$~\msun  & \citep{walter04}    \\
      $M$(CO(7--6), CO(6--5))
                       & $\sim2\times10^{10}$~\msun    & \citep{bertoldi03b}  \\
      $M_{dyn.}$   & $(5.0\pm2.5)\times10^{10}$~\msun
                                                      & \citep{walter04}    \\
      $M_{\rm BH}$     & $3\times10^9$~\msun           & \citep{willott03}   \\
       & & \\
\enddata
\label{j11_obs}
\end{deluxetable}

\begin{deluxetable}{lllll}
\tablewidth{0pt}
\tabletypesize{\footnotesize}
\tablecaption{Dust Properties and IR Luminosities$^1$}
\tablehead{
\colhead { } &
\colhead { } &
\colhead { $T_{dust}$ (K)} &
\colhead { $M_{dust}$ (\msun)} &
\colhead{$L_{\rm IR}$ (\lsun)} 
}
\startdata
Graphite		& &  49 & $2.7\times 10^8$  & $1.9\times 10^{13}$ \\ 
Silicate		& &  47 & $4.9\times 10^8$  & $2.0\times 10^{13} $ \\
Carbon BE	&  & 64 & $9.7\times 10^7$  & $2.4\times 10^{13} $ \\
Carbon AC	&  & 74 & $9.3\times 10^7$  & $2.9\times 10^{13} $ \\
\enddata
\tablenotetext{1}{Graphite and silicate optical properties were taken from \citep{draine84}, and the optical properties of  the carbon dust were taken from \citep{rouleau91}.}
\label{dustprop}
\end{deluxetable}

\begin{deluxetable}{lll}
\tablewidth{0pt}
\tabletypesize{\footnotesize}
\tablecaption{Summary of Derived and Adopted Properties of \jay\ $^1$}
\tablehead{
\colhead { Quantity} &
\colhead { } &
\colhead {Value } 
}
\startdata
Age 		&  & 400~Myr\\ 
Initial Mass (closed box model) 	& &  $5\times10^{10}$~\msun \\ 
Gas mass at 400~Myr 		& &  $3\times10^{10}$~\msun \\ 
IR luminosity		&   &  $2.0\times 10^{13} $~\lsun\\
UV-optical luminosity  &   &  $2.0\times 10^{13} $~\lsun \\
Dust Mass	&   & $2\times 10^8 $~\msun \\
Star formation rate		& &  10--3500~\myr \\
\enddata
\tablenotetext{1}{At $z=6.4$.}
\label{adopt}
\end{deluxetable}


\begin{deluxetable}{lllr}
  \tablecolumns{4}
  \tablewidth{0pc}
  \tablecaption{Observed Fluxes From \jay.}
  \tablehead{
    \colhead{$\lambda_{obs}$} & \colhead{$\lambda_{rest}$}
     & \colhead{$F_{obs}$} & \colhead{References}  \\
     \colhead{(\mic)} & \colhead{(\mic)} & \colhead{(mJy)} & }
  \startdata
        0.77  &      0.10  &    $0.0017\pm0.0002$  & \citep{fan03}           \\
        0.91  &      0.12  &    $0.0325\pm0.0033$  & \citep{fan03}           \\
        1.08  &      0.14  &    $0.0887\pm0.0089$  & \citep{fan03}           \\
        1.22  &      0.16  &    $0.0796\pm0.0080$  & \citep{fan03}           \\
        1.63  &      0.22  &    $0.091\pm0.009$    & \citep{willott03}       \\
        2.19  &      0.30  &    $0.103\pm0.010$    & \citep{willott03}       \\
         3.6  &      0.49  &    $0.124\pm0.002$    & \citep{jiang06}         \\
         4.5  &      0.61  &    $0.140\pm0.003$    & \citep{jiang06}         \\
         5.8  &      0.78  &    $0.133\pm0.010$    & \citep{jiang06}         \\
         8.0  &      1.08  &    $0.241\pm0.016$    & \citep{jiang06}         \\
          16  &      2.16  &    $0.51\pm0.25$      & \citep{charmandaris04}  \\
          22  &      2.96  &    $0.74\pm0.37$      & \citep{charmandaris04}  \\
          24  &      3.23  &    $1.52\pm0.13$      & \citep{jiang06}         \\
          70  &      9.43  &    $\lesssim10$       & \citep{jiang06}         \\
         350  &      47.2  &    $21.0\pm8.1$       & \citep{beelen06}        \\
         450  &      60.7  &    $24.7\pm8.6$       & \citep{robson04}        \\
         850  &       115  &    $7.8\pm0.7$        & \citep{robson04}        \\
        1200  &       162  &    $5.0\pm1.1$        & \citep{bertoldi03a}      \\
        3000  &       404  &    $\lesssim0.52$     & \citep{bertoldi03b}     \\
  $2.1\times 10^{5}$  & $2.9\times 10^{4}$ &    $0.055\pm0.012$    & \citep{carilli04}       \\
  \enddata
  \label{tab:SED}
\end{deluxetable}

\end{document}